\RequirePackage{fix-cm}
\documentclass[twocolumn]{svjour3}          
\smartqed  
\usepackage{graphicx}
\usepackage[numbers,sort&compress]{natbib}  
\usepackage{multirow}
\usepackage{amssymb}
\usepackage{ulem}
\usepackage{xcolor}
\usepackage{threeparttable}  
\PassOptionsToPackage{hyphens}{url}
\usepackage{hyperref}  
\usepackage{array}
\usepackage{amsmath}
\usepackage{makecell}
\usepackage{breqn}
\usepackage[linesnumbered,ruled,vlined]{algorithm2e}
\usepackage{url}
\begin{document}

\title{Survey of Filtered Approximate Nearest Neighbor Search over the Vector-Scalar Hybrid Data}


\author{Yanjun Lin \and
        Kai Zhang \and 
        Zhenying He \and 
        Yinan Jing \and 
        X. Sean Wang 
}


\institute{Yanjun Lin \at
           School of Computer Science, Fudan University, Shanghai, China \\
           \email{yjlin24@m.fudan.edu.cn}           
           \and
           Kai Zhang \at
           School of Computer Science, Fudan University, Shanghai, China \\
           \email{zhangk@fudan.edu.cn}
           \and
           Zhenying He \at
           School of Computer Science, Fudan University, Shanghai, China \\
           \email{zhenying@fudan.edu.cn}
           \and
           Yinan Jing \at
           School of Computer Science, Fudan University, Shanghai, China \\
           \email{jingyn@fudan.edu.cn}
           \and
           X. Sean Wang \at
           School of Computer Science, Fudan University, Shanghai, China \\
           \email{xywangCS@fudan.edu.cn}
}

\date{Received: date / Accepted: date}

\maketitle

\begin{abstract}
Filtered approximate nearest neighbor search (FANNS), an extension of approximate nearest neighbor search (ANNS) that incorporates scalar filters, has been widely applied to constrained retrieval of vector data. Despite its growing importance, no dedicated survey on FANNS over the vector-scalar hybrid data currently exists, and the field has several problems, including inconsistent definitions of the search problem, insufficient framework for algorithm classification, and incomplete analysis of query difficulty. This survey paper formally defines the concepts of hybrid dataset and hybrid query, as well as the corresponding evaluation metrics. Based on these, a pruning-focused framework is proposed to classify and summarize existing algorithms, providing a broader and finer-grained classification framework compared to the existing ones. In addition, a review is conducted on representative hybrid datasets, followed by an analysis on the difficulty of hybrid queries from the perspective of distribution relationships between data and queries. This paper aims to establish a structured foundation for FANNS over the vector-scalar hybrid data, facilitate more meaningful comparisons between FANNS algorithms, and offer practical recommendations for practitioners. The code used for downloading hybrid datasets and analyzing query difficulty is available at \url{https://github.com/lyj-fdu/FANNS}.

\end{abstract}

\section{Introduction}
\label{sec:Introduction}

Nearest neighbor search (NNS) over the vector data has been widely used in various real-world applications, such as embedding-based retrieval (EBR) for search engines and recommendation systems \cite{DBLP:conf/kdd/BorisyukMM0LMBR21, DBLP:conf/kdd/DuRCCWRLHZLT022, DBLP:conf/kdd/LiLJLYZWM21, DBLP:conf/kdd/MagnaniLCYSPC0K22, DBLP:conf/www/HeT0CYTCZKP23, DBLP:conf/cikm/LinYLRSCCMLML24}, retrieval-augmented generation (RAG) for large language models (LLMs) \cite{DBLP:conf/nips/LewisPPPKGKLYR020, DBLP:journals/corr/abs-2312-10997, DBLP:journals/corr/abs-2404-10981, DBLP:journals/corr/abs-2402-19473, DBLP:journals/corr/abs-2404-19543, DBLP:journals/corr/abs-2409-14924}, and many other cross-disciplinary usages \cite{DBLP:conf/ide/AbramovP24, DBLP:conf/clef/BarachanouTP24, DBLP:journals/patterns/KosonockyFWE23, DBLP:conf/aaai/ChangYXH24, DBLP:journals/kbs/GuptaLGD23, DBLP:conf/ecir/KirchoffWHMPGT24}. Typically, unstructured data (e.g., images, texts, or audio) are first encoded into high-dimensional feature vectors using embedding models (e.g., CNN \cite{DBLP:journals/pieee/LeCunBBH98, DBLP:journals/corr/SimonyanZ14a, DBLP:conf/cvpr/HeZRS16}, Transformer \cite{DBLP:conf/nips/VaswaniSPUJGKP17, DBLP:conf/nips/BrownMRSKDNSSAA20, DBLP:conf/naacl/DevlinCLT19}, or VGGish \footnote{\url{https://github.com/tensorflow/models/tree/master/research/audioset/vggish}}), and then nearest neighbor search is performed by retrieving the nearest vectors to a query vector based on a vector similarity function. To alleviate the prohibitively high computational complexity of exact NNS, approximate nearest neighbor search (ANNS) has been proposed to relax the requirement for exactness while significantly improving the search efficiency. ANNS is typically implemented using a well-designed index, which can be hash-based \cite{DBLP:conf/vldb/GionisIM99, DBLP:conf/stoc/Charikar02, DBLP:conf/nips/KulisD09, DBLP:conf/cvpr/HeoLHCY12, DBLP:journals/tcyb/JinLLC14}, tree-based \cite{DBLP:journals/cacm/Bentley75, DBLP:conf/vldb/CiacciaPZ97, DBLP:conf/stoc/DasguptaF08, DBLP:conf/visapp/MujaL09, DBLP:conf/sigir/LiAZM0LC23}, quantization-based \cite{DBLP:conf/cvpr/0002F13, DBLP:journals/pami/JegouDS11, DBLP:conf/eccv/MartinezCHL16, DBLP:conf/eccv/MartinezZHL18, DBLP:journals/sensors/ChenGW10}, or graph-based \cite{DBLP:journals/is/MalkovPLK14, DBLP:conf/nips/SubramanyaDSKK19, DBLP:journals/pami/MalkovY20, DBLP:journals/tdasci/ZhangMSHXGG21, DBLP:journals/pvldb/FuXWC19, DBLP:journals/pami/FuWC22, DBLP:journals/pacmmod/PengCCYX23}.


Filtered nearest neighbor search (FNNS) over the vector-scalar hybrid data extends NNS to retrieving only the nearest vectors among those that satisfy a given scalar filter. It has attracted increasing attention over the past decade. In the context of EBR, users on an e-commerce platform may need to find products most similar to an item in a given image, while applying a filter on scalars like color or price \cite{DBLP:journals/pvldb/WeiWWLZ0C20}. For RAG, in order to ensure the freshness of responses from LLMs, time-aware RAG can be achieved by assigning different weights to document timestamps during retrieval \cite{DBLP:journals/corr/abs-2312-10997}. In the case of industrial-scale knowledge graphs (KGs) such as Saga \cite{DBLP:conf/sigmod/IlyasRKPQS22}, users can find related entities via KG embeddings, while also specifying the entity type and the values they contain \cite{DBLP:journals/pacmmod/MohoneyPCMIMPR23}. In the literature, there has been a surge of studies  \cite{ACM:10.1145/3698822, DBLP:journals/corr/abs-2409-02571, DBLP:conf/icml/EngelsLYDS24, DBLP:journals/pacmmod/ZuoQZLD24, DBLP:journals/pacmmod/PatelKGZ24, DBLP:journals/corr/abs-2308-15014, DBLP:journals/pacmmod/MohoneyPCMIMPR23, DBLP:conf/www/GollapudiKSKBRL23, DBLP:conf/osdi/ZhangXCSXCCH00Y23, DBLP:conf/cikm/WuHQFLY22, DBLP:conf/nips/WangLX0YN23, DBLP:journals/corr/abs-2210-14958, DBLP:conf/sigmod/WangYGJXLWGLXYY21, DBLP:journals/pvldb/WeiWWLZ0C20, DBLP:conf/sigmod/YangLFW20, DBLP:journals/concurrency/XuLWX20} focused on solving the problem of filtered approximate nearest neighbor search (FANNS) over the vector-scalar hybrid data.

\subsection{Motivation}
\label{subsec:Motivation}

While a number of survey papers have been published on ANNS over the vector data \cite{DBLP:journals/tkde/LiZSWLZL20, DBLP:journals/is/AumullerBF20, DBLP:journals/tkde/Cai21, DBLP:journals/pvldb/WangXY021, DBLP:journals/debu/00070P023, DBLP:journals/vldb/PanWL24}, there is currently no survey on FANNS over the vector-scalar hybrid data. Below, we outline three key reasons (\textbf{R1}-\textbf{R3}) that highlight the need for such a survey.


\paragraph{R1: Inconsistent definitions of the search problem.} In FANNS, datasets and queries containing both vectors and scalars are referred as hybrid datasets and hybrid queries, respectively. However, the definitions of these terms vary across studies. For hybrid datasets, some define scalars as simple numbers \cite{DBLP:journals/corr/abs-2308-15014, DBLP:journals/pacmmod/ZuoQZLD24, DBLP:conf/icml/EngelsLYDS24, DBLP:journals/corr/abs-2409-02571}, some as collections of labels \cite{DBLP:conf/www/GollapudiKSKBRL23, ACM:10.1145/3698822}, and others as values within a schema \cite{DBLP:journals/concurrency/XuLWX20, DBLP:conf/sigmod/YangLFW20, DBLP:journals/pvldb/WeiWWLZ0C20, DBLP:conf/sigmod/WangYGJXLWGLXYY21, DBLP:journals/corr/abs-2210-14958, DBLP:conf/nips/WangLX0YN23, DBLP:conf/cikm/WuHQFLY22, DBLP:conf/osdi/ZhangXCSXCCH00Y23, DBLP:journals/pacmmod/MohoneyPCMIMPR23, DBLP:journals/pacmmod/PatelKGZ24} similar to that in a relational database. For hybrid queries, some define scalar filters as equality comparisons \cite{DBLP:journals/concurrency/XuLWX20, DBLP:conf/nips/WangLX0YN23, DBLP:conf/cikm/WuHQFLY22, DBLP:conf/www/GollapudiKSKBRL23, ACM:10.1145/3698822}, some as range comparisons \cite{DBLP:journals/pacmmod/ZuoQZLD24, DBLP:conf/icml/EngelsLYDS24, DBLP:journals/corr/abs-2409-02571}, and others as general operations \cite{DBLP:conf/sigmod/YangLFW20, DBLP:journals/pvldb/WeiWWLZ0C20, DBLP:conf/sigmod/WangYGJXLWGLXYY21, DBLP:journals/corr/abs-2210-14958, DBLP:conf/osdi/ZhangXCSXCCH00Y23, DBLP:journals/pacmmod/MohoneyPCMIMPR23, DBLP:journals/corr/abs-2308-15014, DBLP:journals/pacmmod/PatelKGZ24}. Additionally, there is inconsistency in how evaluation metrics are defined. For example, some studies define the \textit{selectivity} as the proportion of data points that do not satisfy the scalar filter  \cite{DBLP:conf/sigmod/WangYGJXLWGLXYY21, DBLP:journals/pvldb/WeiWWLZ0C20}, while others define it as the proportion of data points that do \cite{DBLP:journals/pacmmod/MohoneyPCMIMPR23, DBLP:journals/pacmmod/PatelKGZ24, DBLP:conf/osdi/ZhangXCSXCCH00Y23}. The latter definition is also referred to as the \textit{specificity} in some studies \cite{DBLP:conf/www/GollapudiKSKBRL23, ACM:10.1145/3698822}.

\paragraph{R2: Insufficient framework for algorithm classification.} Current framework classifies FANNS algorithms based on \textit{when} the scalar filter is applied, typically into three categories: pre-filtering, post-filtering, and in-filtering \cite{DBLP:journals/corr/abs-2409-02571, DBLP:conf/icml/EngelsLYDS24, DBLP:journals/pacmmod/PatelKGZ24, DBLP:journals/corr/abs-2308-15014, DBLP:conf/www/GollapudiKSKBRL23, DBLP:journals/pacmmod/MohoneyPCMIMPR23}, which correspond to removing data points that do not satisfy the filter before, after, or during the ANNS. However, this framework has two major shortcomings. First, it is not enough to cover all algorithms. For example, some algorithms \cite{DBLP:conf/icml/EngelsLYDS24, DBLP:journals/pacmmod/MohoneyPCMIMPR23} first apply part of the scalar filter to identify relevant data partitions, then perform ANNS within them, and finally apply the complete filter. In this case, the filter is applied in two stages, which fails to fit any of the aforementioned three categories. Second, it is too coarse to distinguish between algorithms. For example, in graph-based indices, the filter can be applied during either result update \cite{DBLP:conf/osdi/ZhangXCSXCCH00Y23} or neighborhood expansion \cite{DBLP:journals/pacmmod/PatelKGZ24}, but only the latter avoids unnecessary vector similarity calculations. Although both are classified as in-filtering, their effects differ significantly. 

\paragraph{R3: Incomplete analysis of query difficulty.} In the context of ANNS, considerable efforts have been made to understand query difficulty through factors such as \textit{relative contrast} \cite{DBLP:conf/icml/HeKC12, DBLP:journals/is/AumullerC21}, \textit{intrinsic dimensionality} \cite{DBLP:conf/icdm/Houle13, DBLP:conf/sisap/0001C19, DBLP:journals/is/AumullerC21}, \textit{query expansion} \cite{DBLP:conf/soda/AhleAP17, DBLP:journals/is/AumullerC21}, \textit{$\epsilon$-hardness} \cite{DBLP:journals/vldb/ZoumpatianosLIP18}, and \textit{Steiner-hardness} \cite{DBLP:journals/corr/abs-2408-13899}. In contrast, for FANNS, factors that impact the difficulty of hybrid queries remain underexplored. At present, \textit{selectivity} is the only commonly used factor for evaluating the hybrid query difficulty. Identifying more factors is essential not only for explaining algorithm performance fluctuations from various perspectives, but also for constructing evaluation benchmarks across a wider range of difficulty levels by their combination.

\subsection{Our Contributions}
\label{subsec:Our Contributions}

Driven by the above discussion, we present a survey on FANNS over the vector-scalar hybrid data, covering its definitions, algorithms, datasets and query difficulty. Below, we summarize our main contributions.

\paragraph{(1) More systematic definition of the search problem.} For \textbf{R1}, we formally define the problem of FANNS by specifying the hybrid dataset, the hybrid query, and relevant evaluation metrics (\autoref{sec:Preliminaries}). Specifically, a hybrid dataset is defined as one in which each data point is a pair of a scalar-tuple following a scalar schema and a vector in a vector space. A hybrid query is defined as comprising a scalar filter, a vector similarity function, a query vector, and a target result size. Key evaluation metrics including \textit{recall} and \textit{selectivity} are defined with precise semantics to eliminate ambiguity.

\paragraph{(2) Finer-grained classification of FANNS algorithms.} For \textbf{R2}, we propose a pruning-focused framework to classify FANNS algorithms (\autoref{sec:Review of FANNS Algorithms}). Our framework comprises four distinct strategies, each emphasizing either vector pruning or scalar pruning: vector-solely pruning (VSP), vector-centric joint pruning (VJP), scalar-centric joint pruning (SJP), and scalar-solely pruning (SSP). Building on this framework, we summarize existing FANNS algorithms using the unified terminology in our formal definitions, effectively classifying them and revealing their interrelationships. Our framework provides a broader and finer-grained classification compared to the existing one.

\paragraph{(3) Deeper analysis of query difficulty through a new factor.} For \textbf{R3}, we examine existing hybrid datasets (\autoref{sec:Review of Hybrid Datasets}) and analyze the factors that impact the difficulty of hybrid queries (\autoref{sec:The Distribution Factor for Query Difficulty}). Specifically, we collect hybrid datasets used in existing FANNS studies, discuss their underlying construction strategies, and detail their main characteristics. Motivated by realistic scenarios, we identify the \textit{distribution} factor, which refers to the relationship between high dimensional distributions of two sets of vectors. We verify the impact of the \textit{distribution} factor to the hybrid query difficulty by conducting a case study, offering qualitative explanations based on UMAP visualizations \cite{DBLP:journals/corr/abs-1802-03426} and quantitative insights using the Wasserstein distance \cite{Mahalanobis1936}. We finally propose a schema towards more comprehensive evaluation of FANNS algorithms combining both \textit{selectivity} and \textit{distribution} factors.

\section{Preliminaries}
\label{sec:Preliminaries}

In this section, we first formally define the hybrid dataset, the hybrid query, and the corresponding evaluation metrics. Then, we outline representative ANNS algorithms, which serve as the foundation for FANNS algorithms. \autoref{tab:notations} summarizes the notations used in this paper.

\subsection{Definition of Hybrid Dataset}
\label{subsec:Definition of Hybrid Dataset}

We begin with the definitions of scalar schema and vector space, followed by a formal definition of hybrid dataset.

\begin{definition}[Scalar Schema]
Let $\mathbb{S} = (\mathbb{S}_{1}, \mathbb{S}_{2}, \dots, \newline \mathbb{S}_{m})$ be a \textit{scalar schema} containing $m$ \textit{scalars}, where each scalar $\mathbb{S}_{i}$ has a simple data type, such as integer, float, or string.
\end{definition}

\begin{definition}[Vector Space]
Let $\mathbb{V}$ be a \textit{vector space}, which is a set of vectors that are closed under vector addition and constant multiplication. In this paper, we specifically consider the $d$-dimensional vector space over the field of real numbers, i.e., $\mathbb{V} = \mathbb{R}^{d}$, where each vector is a $d$-dimensional tuple of real numbers.
\end{definition}

\begin{definition}[Hybrid Dataset]
Let $\mathcal{D} = \{\mathbf{p}_{1}, \mathbf{p}_{2}, \newline \dots, \mathbf{p}_{n}\} = \{(\mathbf{s}_{1}, \mathbf{v}_{1}), (\mathbf{s}_{2}, \mathbf{v}_{2}), \dots, (\mathbf{s}_{n}, \mathbf{v}_{n})\}$ be a \textit{hybrid dataset} containing $n$ data points, where each \textit{data point} $\mathbf{p}_{i} = (\mathbf{s}_{i}, \mathbf{v}_{i})$ is a pair of a \textit{scalar-tuple} $\mathbf{s}_{i} \in \mathbb{S}$ (also denoted as $\mathbf{p}_{i}.\mathbf{s}$) and a \textit{vector} $\mathbf{v}_{i} \in \mathbb{R}^{d}$ (also denoted as $\mathbf{p}_{i}.\mathbf{v}$). Each scalar-tuple $\mathbf{s}_{i} = (s_{i,1}, s_{i,2}, \dots, s_{i,m})$ is a value in $\mathbb{S}$, where each $s_{i,j}$ is a \textit{scalar value} in $\mathbb{S}_{i}$ and can either take a value of the corresponding data type or be NULL. Each vector $\mathbf{v}_{i} = (v_{i,1}, v_{i,2}, \dots, v_{i,d})$ is a value in $\mathbb{R}^{d}$, where each $v_{i,j}$ is a real number in $\mathbb{R}$.
\end{definition}

\begin{table}[tbp]
  \caption{Summary of Notations}
  \label{tab:notations}
  \renewcommand{\arraystretch}{1.1}
  \centering
  \begin{tabular}{l|p{0.36\textwidth}} 
    \hline
    Notation & Description \\
    \hline
    \noalign{\vskip 2pt}
    \hline

    $\mathbb{S}$ & A scalar schema, consisting of $m$ scalars, i.e., $\mathbb{S}=(\mathbb{S}_{1}, \mathbb{S}_{2}, \dots, \mathbb{S}_{m})$, where each scalar $\mathbb{S}_{j}$ has a simple data type \\
    \hline
    $\mathbb{V}$ & A vector space, specifically the $d$-dimensional vector space over the field of real numbers, i.e., $\mathbb{V} = \mathbb{R}^{d}$ \\
    \hline
    $\mathcal{D}$ & A hybrid dataset, consisting of $n$ data points, i.e., $\mathcal{D}=\{\mathbf{p}_{1}, \mathbf{p}_{2}, \dots, \mathbf{p}_{n}\}$, where each data point $\mathbf{p}_{i} = (\mathbf{s}_{i}, \mathbf{v}_{i})$ has a scalar-tuple $\mathbf{s}_{i} \in \mathbb{S}$ and a vector $\mathbf{v}_{i} \in \mathbb{R}^{d}$ \\
    \hline
    \noalign{\vskip 2pt}
    \hline
    
    $f_{s}$ & A scalar filter, $f_{s}:\mathbb{S} \rightarrow \{0,1\}$, which evaluates a scalar-tuple in $\mathbb{S}$ to false or true \\
    \hline
    $\mathcal{D}_{f_{s}}$ & A filtered subset, $\mathcal{D}_{f_{s}} = \{\mathbf{p} \in \mathcal{D} \mid f_{s}(\mathbf{p}.\mathbf{s}) = 1\}$, which contains data points that satisfy $f_{s}$ in $\mathcal{D}$ \\
    \hline
    $f_{v}$ & A vector similarity function, $f_{v}:\mathbb{R}^{d} \times \mathbb{R}^{d} \rightarrow \mathbb{R}$, which measures the similarity between two vectors in $\mathbb{R}^{d}$, where smaller values indicate higher similarity \\ 
    \hline
    $q$ & A hybrid query, $q=(f_{s}, f_{v}, \mathbf{v}_{q}, k)$, containing a scalar filter $f_{s}$, a vector similarity function $f_{v}$, a query vector $\mathbf{v}_{q}$, and target result size $k$ \\ 
    \hline
    \noalign{\vskip 2pt}
    \hline

    $|\cdot|$ & The cardinality of a set \\
    \hline
    $recall@k$ & The recall to measure the accuracy of a hybrid query with a target result size $k$, defined as $recall = \frac{|\mathcal{R} \cap \tilde{\mathcal{R}}|}{|\mathcal{R}|}$, where $\mathcal{R}$ and $\tilde{\mathcal{R}}$ are the result for FNNS and FANNS, respectively \\ 
    \hline
    $sel_{f_{s}}$ & The selectivity of a scalar filter $f_{s}$, defined as $sel_{f_{s}} = 1 - \frac{|\mathcal{D}_{f_{s}}|}{|\mathcal{D}|}$, which is the fraction of data points that do not satisfy the scalar filter $f_{s}$. \\
    \hline
  \end{tabular}
\end{table}

\subsection{Definition of Hybrid Query}
\label{subsec:Definition of Hybrid Query}

We proceed to define the corresponding scalar filter and vector similarity function, along with a formal definition of hybrid query.

\begin{definition}[Scalar Filter]
Let $f_{s}:\mathbb{S} \rightarrow \{0,1\}$ be a \textit{scalar filter} that evaluates the scalar-tuple of a data point $\mathbf{p}$ to either false ($f_{s}(\mathbf{p}.\mathbf{s})=0$) or true ($f_{s}(\mathbf{p}.\mathbf{s})=1$). $\mathcal{D}_{f_{s}} = \{\mathbf{p} \in \mathcal{D} \mid f_{s}(\mathbf{p}.\mathbf{s})=1\}$ is the \textit{filtered subset} whose data points are taken from $\mathcal{D}$ and \textit{satisfy} $f_{s}$.
\end{definition}

A scalar filter can vary in complexity, ranging from simple constraints like binary comparisons to more complex constraints like regular expressions. A scalar filter that allows arbitrary constraints is referred to as a \textit{general scalar filter}. Unless otherwise specified, a scalar filter is assumed to be a general scalar filter.

Some algorithms \cite{DBLP:conf/nips/WangLX0YN23, DBLP:conf/cikm/WuHQFLY22, DBLP:conf/www/GollapudiKSKBRL23, DBLP:journals/pacmmod/ZuoQZLD24, DBLP:journals/corr/abs-2409-02571, DBLP:journals/concurrency/XuLWX20, DBLP:conf/icml/EngelsLYDS24}, discussed later, simplify the scalar filter for specialized applications. A \textit{simplified scalar filter} consists of simple constraints that check whether a scalar equals a specific value (\textit{equality constraints}) or falls within a value range (\textit{range constraints}). A \textit{simplified scalar filter} with only \textit{equality constraints} is called a \textit{simplified equality scalar filter}, while one with only \textit{range constraints} is called a \textit{simplified range scalar filter}. For ease of analysis, we express a \textit{simplified scalar filter} in the disjunctive normal form (DNF):
\begin{equation}
\label{eq:simplified scalar filter}
f_{s}(\mathbf{p}.\mathbf{s}) = \bigvee_{i=1}^{t}\left(\bigwedge_{j=1}^{m} f_{i,j}(\mathbf{p}.s_{j})\right) \in \{0,1\},
\end{equation}
where the $j$-th \textit{sub-filter} $f_{i,j}:\mathbb{S}_{j} \rightarrow \{0,1\}$ imposes either no constraint or a simple constraint on the $j$-th scalar $\mathbf{p}.s_{j} \in \mathbb{S}_{j}$. A sub-filter always evaluates to true when no constraint is imposed, and is referred to as an \textit{active sub-filter} when a simple constraint is specified. 

\begin{definition}[Vector Similarity Function]
Let $f_{v}:\mathbb{R}^{d} \times \mathbb{R}^{d} \rightarrow \mathbb{R}$ be a \textit{vector similarity function} that measures the similarity between the vectors of two data points $\mathbf{p}_{x}$ and $\mathbf{p}_{y}$ to a real number $f_{v}(\mathbf{p}_{x}.\mathbf{v}, \mathbf{p}_{y}.\mathbf{v}) \in \mathbb{R}$, where smaller values indicate higher similarity.
\end{definition}

Given two vectors $\mathbf{p}_{x}.\mathbf{v} = (v_{x,1}, v_{x,2}, \dots, v_{x,d})$ and $\mathbf{p}_{y}.\mathbf{v} = (v_{y,1}, v_{y,2}, \dots, v_{y,d})$ in $\mathbb{R}^{d}$, the form of a vector similarity function $f_{v}$ can vary depending on the specific application. 

A common approach is to directly define the vector similarity function using a distance function, which obeys the properties of non-negativity, identity, symmetry, and the triangle inequality. For example, the Euclidean distance:
\begin{equation}
\label{eq:Euclidean distance}
d(\mathbf{p}_{x}.\mathbf{v}, \mathbf{p}_{y}.\mathbf{v}) = \sqrt{\sum_{i=1}^{d}{(v_{x,i}-v_{y,i})^{2}}} \in [0,+\infty),
\end{equation}  
can serve directly as a vector similarity function, i.e., $f_{v}(\cdot,\cdot) = d(\cdot,\cdot)$. A smaller Euclidean distance value indicates that two vectors are closer in the vector space, implying higher similarity. 

Alternatively, a non-distance function can be used to indirectly to define the vector similarity function. For example, the inner product:
\begin{equation}
\label{eq:inner product}
\langle\mathbf{p}_{x}.\mathbf{v}, \mathbf{p}_{y}.\mathbf{v}\rangle = \sum_{i=1}^{d}{v_{x,i}v_{y,i}} \in \mathbb{R},
\end{equation}  
can be transformed into a vector similarity function by taking its negation, i.e., $f_{v}(\cdot,\cdot) = -\langle\cdot,\cdot\rangle$. A larger inner product value (smaller in negative) indicates that two vectors are more aligned in direction, implying higher similarity.

\begin{definition}[Hybrid Query]
Let $q=\{f_{s}, f_{v}, \mathbf{v}_{q}, k\}$ be a \textit{hybrid query}, which contains a scalar filter $f_{s}$, a vector similarity function $f_{v}$, a \textit{query vector} $\mathbf{v}_{q} \in \mathbb{R}^{d}$, and a \textit{target result size} $k$.
\end{definition}

Given a hybrid dataset $\mathcal{D}$ and a hybrid query $q$, the problems of FNNS is to find a result set $\mathcal{R} \subseteq \mathcal{D}_{f_{s}}$ containing $\min(k,|\mathcal{D}_{f_{s}}|)$ data points whose vectors are the most similar to $\mathbf{v}_{q}$ under $f_{v}$ when only considering vectors in $\mathcal{D}_{f_{s}}$. This can be formally expressed as:
\begin{equation}
\mathcal{R} = \arg \min_{\substack{\mathcal{S} \subseteq \mathcal{D}_{f_{s}}, \\|\mathcal{S}| = \min(k,|\mathcal{D}_{f_{s}}|)}}{{\sum}_{\mathbf{p} \in \mathcal{S}}{f_{v}(\mathbf{p}.\mathbf{v}, \mathbf{v}_{q})}}.
\end{equation}

FANNS relaxes FNNS by allowing a small number of errors in the result set, denoted $\tilde{\mathcal{R}}$, thereby trading accuracy for efficiency. The accuracy of FANNS is typically measured by recall, as defined in \autoref{subsec:Definitions of Evaluation Metrics}. It is worth noting that when the scalar filter $f_{s}$ always evaluates to true, FNNS and FANNS reduce to traditional NNS and ANNS, respectively.


\subsection{Definitions of Evaluation Metrics}
\label{subsec:Definitions of Evaluation Metrics}

This subsection clarifies the semantics of key evaluation metrics, including \textit{recall} and \textit{selectivity}. 

\begin{definition}[Recall]
Let $recall@k$ be the \textit{recall} that measures the accuracy of a hybrid query with a target result size $k$, which is defined as follows:
\begin{equation}
\label{eq:recall}
recall@k = \frac{|\mathcal{R} \cap \tilde{\mathcal{R}}|}{|\mathcal{R}|} \in [0,1].
\end{equation}
\end{definition}

In this equation, $\mathcal{R}$ denotes the ground truth result set obtained from FNNS, while $\tilde{\mathcal{R}}$ represents the result returned by a given query method, which may correspond to either FNNS or FANNS. The numerator, $|\mathcal{R} \cap \tilde{\mathcal{R}}|$, indicates the number of ground truth results that are successfully retrieved. The denominator, $|\mathcal{R}|$, reflects the total number of ground truth results. Note that $|\mathcal{R}| \leq k$, as the \textit{filtered subset} $\mathcal{D}_{f_{s}}$ may contain fewer than $k$ data points.

A larger \textit{recall} indicates higher quality of the result. Specifically, the \textit{recall} of FNNS is $1$, and the \textit{recall} of FANNS is between $0$ and $1$ due to its approximate nature. If multiple vectors have the same similarity to $\mathbf{v}_{q}$ under $f_{v}$, the result set may not be unique, in this case, \textit{distance-based recall} variants are available \cite{DBLP:journals/is/AumullerBF20}.

\begin{definition}[Selectivity]
Let $sel_{f_{s}}$ be the \textit{selectivity} that measures the fraction of data points that do \textbf{not} satisfy the scalar filter $f_{s}$, which is defined as follows:
\begin{equation}
\label{eq:selectivity}
sel_{f_{s}} = 1 - \frac{|\mathcal{D}_{f_{s}}|}{|\mathcal{D}|} \in [0,1].
\end{equation}
\end{definition}

In the existing literature, the definition of \textit{selectivity} remains inconsistent. Some studies define it as the proportion of data points that do \textbf{not} satisfy the scalar filter \cite{DBLP:conf/sigmod/WangYGJXLWGLXYY21, DBLP:journals/pvldb/WeiWWLZ0C20}, while others define it as the proportion of data points that satisfy the filter \cite{DBLP:journals/pacmmod/MohoneyPCMIMPR23, DBLP:journals/pacmmod/PatelKGZ24, DBLP:conf/osdi/ZhangXCSXCCH00Y23}. In this work, we adopt the former definition, where higher selectivity indicates that a larger portion of the dataset is filtered out. This aligns with the intuitive understanding that a process is ``more selective'' when it excludes more candidates. The latter definition is more appropriately referred to as \textit{specificity} \cite{DBLP:conf/www/GollapudiKSKBRL23, ACM:10.1145/3698822}.

A scalar filter $f_{s}$ is said to be \textit{selective} if $sel_{f_{s}}$ is close to $1$, \textit{unselective} if it is close to $0$, and \textit{moderate} if it lies in between. 

\subsection{Background of ANNS Algorithms}
\label{subsec:Background of ANNS Algorithms}


Early research on ANNS primarily focused on hash-based \cite{DBLP:conf/vldb/GionisIM99, DBLP:conf/stoc/Charikar02, DBLP:conf/nips/KulisD09, DBLP:conf/cvpr/HeoLHCY12, DBLP:journals/tcyb/JinLLC14} and tree-based \cite{DBLP:journals/cacm/Bentley75, DBLP:conf/vldb/CiacciaPZ97, DBLP:conf/stoc/DasguptaF08, DBLP:conf/visapp/MujaL09, DBLP:conf/sigir/LiAZM0LC23} indices, as they are natural extensions of traditional indexing structures in relational databases to high-dimensional spaces. However, despite their favorable theoretical complexity, these methods fail to scale effectively when the vector dimensionality exceeds $10$ \cite{DBLP:journals/corr/abs-2401-08281}, largely due to the ``curse of dimensionality'' \cite{DBLP:conf/stoc/IndykM98}.

In recent years, the focus of ANNS has shifted toward quantization-based \cite{DBLP:conf/cvpr/0002F13, DBLP:journals/pami/JegouDS11, DBLP:conf/eccv/MartinezCHL16, DBLP:conf/eccv/MartinezZHL18, DBLP:journals/sensors/ChenGW10} and graph-based \cite{DBLP:journals/is/MalkovPLK14, DBLP:conf/nips/SubramanyaDSKK19, DBLP:journals/pami/MalkovY20, DBLP:journals/tdasci/ZhangMSHXGG21, DBLP:journals/pvldb/FuXWC19, DBLP:journals/pami/FuWC22, DBLP:journals/pacmmod/PengCCYX23} indices, which have demonstrated significantly better empirical performance under different efficiency-accuracy trade-offs \cite{DBLP:journals/corr/abs-2401-08281}. As a result, almost all existing FANNS algorithms are built upon adaptations of these two index structures. Therefore, we briefly introduce the IVF index and the graph index as representative ANNS methods to facilitate the subsequent discussion of the FANNS algorithms.


\begin{algorithm}[tbp]
\caption{Search on the IVF Index}
\label{alg:Search on the IVF Index}

\KwIn{$nlist$ centroids $\mathcal{C} = \{\mathbf{v}_{c_{1}}, \dots, \mathbf{v}_{c_{nlist}}\}$ with inverted lists $\mathcal{L} = \{l_{c_{1}}, \dots, l_{c_{nlist}}\}$, target number of visited lists $nprobe$, vector similarity function $f_v$, query vector $\mathbf{v}_q$, target result size $k$}
\KwOut{approximate query result $\tilde{\mathcal{R}}$}

$\tilde{\mathcal{R}} \leftarrow \{\}$ \tcp*{Result set}
$\mathcal{C'} \leftarrow \{nprobe$ centroids closest to $\mathbf{v}_q$ in $\mathcal{C}\}$\; \label{alg:line:nprobe}
\ForEach{centroid $\mathbf{v}_c \in \mathcal{C'}$}{ \label{alg:line:scan beg}
    $l_c \leftarrow$ inverted list of $\mathbf{v}_c$ from $\mathcal{L}$\;
    \ForEach{vector $\mathbf{v} \in l_c$}{ \label{alg:line:scan end}
        $\mathbf{v}_{f} \leftarrow \arg \max_{\mathbf{v}' \in \tilde{\mathcal{R}}} f_v(\mathbf{v}', \mathbf{v}_q)$\; \label{alg:line:sort beg}
        \If{$|\tilde{\mathcal{R}}| < k$ \textbf{or} $f_v(\mathbf{v}, \mathbf{v}_q) < f_v(\mathbf{v}_f, \mathbf{v}_q)$}{
            $\tilde{\mathcal{R}} \leftarrow \tilde{\mathcal{R}} \cup \{\mathbf{v}\}$\;
            \If{$|\tilde{\mathcal{R}}| > k$}{
                $\mathbf{v}_{f} \leftarrow \arg \max_{\mathbf{v}' \in \tilde{\mathcal{R}}} f_v(\mathbf{v}', \mathbf{v}_q)$\;
                $\tilde{\mathcal{R}} \leftarrow \tilde{\mathcal{R}} \setminus \{\mathbf{v}_f\}$\; \label{alg:line:sort end}
            }
        }
    }
}
\Return $\tilde{\mathcal{R}}$\;
\end{algorithm}

\begin{algorithm}[tbp]
\caption{Search on the Graph Index}
\label{alg:Search on the Graph Index}

\KwIn{graph $G(\mathcal{V}, \mathcal{E})$, entry vector $\mathbf{v}_e$, vector similarity function $f_{v}$, query vector $\mathbf{v}_q$, target result size $k$}
\KwOut{approximate query result $\tilde{\mathcal{R}}$}

$\tilde{\mathcal{R}} \leftarrow \{\mathbf{v}_e\}$ \tcp*{Result set}
$\mathcal{C} \leftarrow \{\mathbf{v}_e\}$ \tcp*{Candidate set}  
\While{$|\mathcal{C}| > 0$}{
    $\mathbf{v}_c \leftarrow \arg \min_{\mathbf{v}' \in \mathcal{C}} f_{v}(\mathbf{v}', \mathbf{v}_q)$\; \label{alg:line:neighborhood expansion beg} 
    $\mathcal{C} \leftarrow \mathcal{C} \setminus \{\mathbf{v}_c\}$\;
    \ForEach{$\mathbf{v} \in neighborhood(\mathbf{v}_c)$}{ \label{alg:line:neighborhood expansion end}
        $\mathbf{v}_{f} \leftarrow \arg \max_{\mathbf{v}' \in \tilde{\mathcal{R}}} f_{v}(\mathbf{v}', \mathbf{v}_q)$\; \label{alg:line:result update beg}
        \If{$|\tilde{\mathcal{R}}| < k$ \textbf{or} $f_{v}(\mathbf{v}, \mathbf{v}_q) < f_{v}(\mathbf{v}_{f}, \mathbf{v}_q)$}{
            $\mathcal{C} \leftarrow \mathcal{C} \cup \{\mathbf{v}\}$\;
            $\tilde{\mathcal{R}} \leftarrow \tilde{\mathcal{R}} \cup \{\mathbf{v}\}$\;
            \If{$|\tilde{\mathcal{R}}| > k$}{
                $\mathbf{v}_{f} \leftarrow \arg \max_{\mathbf{v}' \in \tilde{\mathcal{R}}} f_{v}(\mathbf{v}', \mathbf{v}_q)$\;
                $\tilde{\mathcal{R}} \leftarrow \tilde{\mathcal{R}} \setminus \{\mathbf{v}_f\}$\; \label{alg:line:result update end}
            }
        }
    }
}
\Return $\tilde{\mathcal{R}}$\;
\end{algorithm}

\paragraph{IVF index.} The inverted file (IVF) index is a form of quantization-based indices. It first trains a quantizer, typically using k-means clustering \cite{DBLP:conf/cvpr/0002F13}, to partition the vector space $\mathbb{R}^{d}$ into $nlist$ clusters represented by their centroids, denoted $\mathcal{C} = \{\mathbf{v}_{c_{1}}, \mathbf{v}_{c_{2}}, \dots, \mathbf{v}_{c_{nlist}}\}$, where each centroid $\mathbf{v}_{c_{i}}$ is a value in $\mathbb{R}^{d}$. Each vector in the dataset is then assigned to its nearest centroid, and an inverted list is created for each cluster to store references to the vectors within it. Together, the centroids $\mathcal{C}$ and their corresponding inverted lists $\mathcal{L}$ form the IVF index. For more details, please refer to \cite{DBLP:journals/corr/abs-2401-08281}.

During the search (\autoref{alg:Search on the IVF Index}), the traversal is performed within the $nprobe$ clusters whose centroids are closest to the query vector (line~\ref{alg:line:nprobe}), and all the vectors within these selected clusters are scanned (lines~\ref{alg:line:scan beg}-~\ref{alg:line:scan end}) to find the $k$ nearest neighbors of the query vector $\mathbf{v}_{q}$ (lines~\ref{alg:line:sort beg}-~\ref{alg:line:sort end}). Optionally, compression techniques such as product quantization \cite{DBLP:journals/pami/JegouDS11} or additive quantization \cite{DBLP:journals/sensors/ChenGW10, DBLP:conf/eccv/MartinezCHL16, DBLP:conf/eccv/MartinezZHL18} can be applied to vectors to save memory and accelerate the search process.

\paragraph{Graph Index.} In a graph index, each vector is represented as a vertex, with edges linking vertices whose vectors are similar. Together, the vertex set $\mathcal{V}$ and the edge set $\mathcal{E}$ form the graph index $G(\mathcal{V}, \mathcal{E})$. The index construction typically follows one of the three strategies: incremental construction, refinement, or divide-and-conquer. For more details, please refer to \cite{DBLP:journals/pvldb/WangXY021}.

During search (\autoref{alg:Search on the Graph Index}), the traversal follows a greedy routing strategy, beginning with an entry vector $\mathbf{v}_{e}$ and gradually expanding the neighbors towards the query vector $\mathbf{v}_{q}$. This process involves repeated steps of \textit{neighborhood expansion} (lines~\ref{alg:line:neighborhood expansion beg}-~\ref{alg:line:neighborhood expansion end}) and \textit{result update} (lines~\ref{alg:line:result update beg}-~\ref{alg:line:result update end}). Optionally, several optimizations can be applied to enhance the search process, such as maintaining a visited set \cite{DBLP:journals/is/MalkovPLK14, DBLP:conf/nips/SubramanyaDSKK19, DBLP:journals/pvldb/FuXWC19, DBLP:journals/pami/MalkovY20, DBLP:journals/tdasci/ZhangMSHXGG21, DBLP:journals/pami/FuWC22} to avoid redundant calculations, expanding the result set to include more than $k$ vectors and retaining only the top-$k$ upon returning \cite{DBLP:journals/is/MalkovPLK14, DBLP:conf/nips/SubramanyaDSKK19, DBLP:journals/pvldb/FuXWC19, DBLP:journals/pami/MalkovY20, DBLP:journals/tdasci/ZhangMSHXGG21, DBLP:journals/pami/FuWC22} to improve the accuracy, and utilizing hierarchical graph structures \cite{DBLP:journals/pami/MalkovY20, DBLP:journals/tdasci/ZhangMSHXGG21} to reduce search time.

\section{Review of FANNS Algorithms}
\label{sec:Review of FANNS Algorithms}

\begin{figure*}[tbp]
  \centering
  \includegraphics[width=\linewidth]{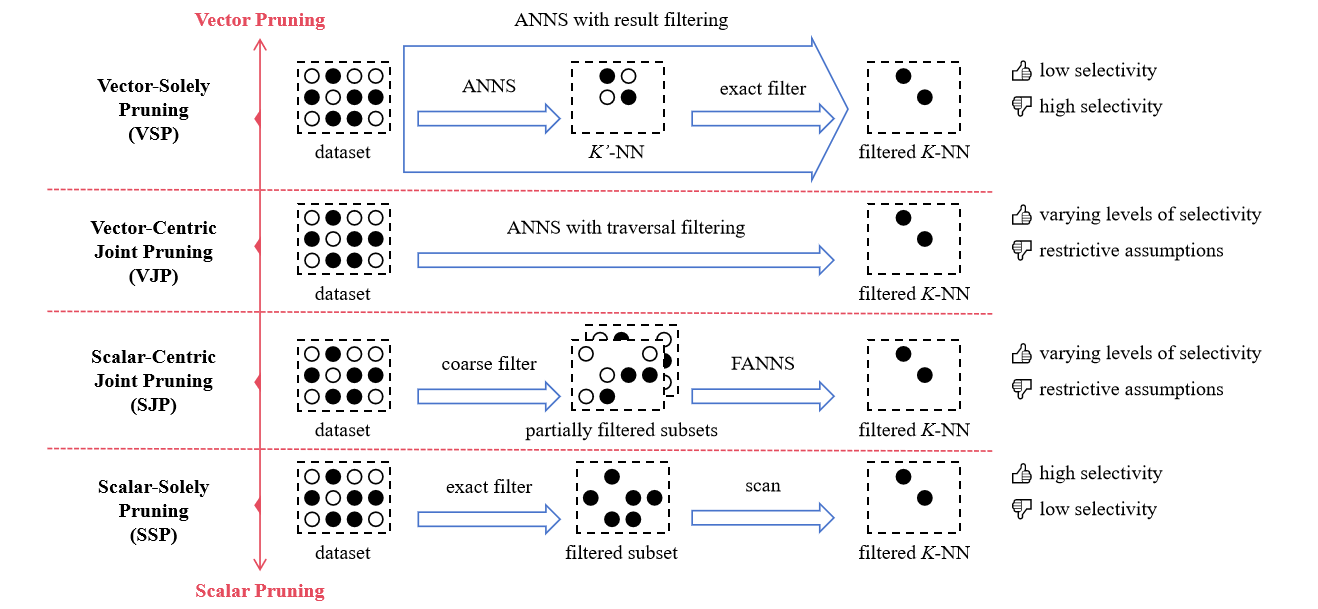}
  \caption{The proposed pruning-focused framework for classifying FANNS algorithms}
  \label{fig:framework}
\end{figure*}
\begin{figure*}[tbp]
  \centering
  \includegraphics[width=\linewidth]{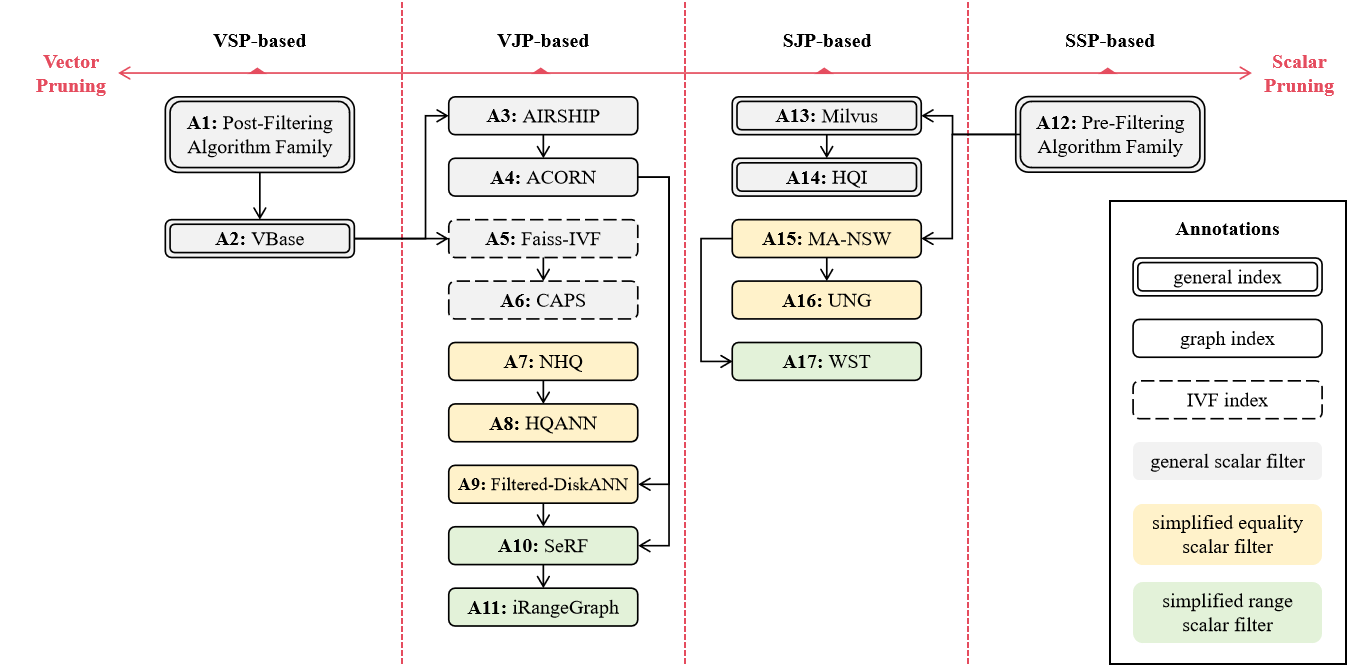}
  \caption{Classification of FANNS algorithms under the pruning-focused framework and interrelationships among them.}
  \label{fig:roadmap}
\end{figure*}

In this section, we propose a pruning-focused framework (\autoref{fig:framework}) to classify 17 existing FANNS algorithms (\textbf{A1}–\textbf{A17}), demonstrating how each algorithm aligns with one of the four distinct pruning strategies defined in our framework and revealing the interrelationships among these algorithms (\autoref{fig:roadmap}). The core design of each algorithm is summarized using the terminology given in \autoref{sec:Preliminaries}.

\subsection{The Pruning-Focused Framework}
\label{subsec:The Pruning-Focused Framework}

Existing framework classifies FANNS algorithms based on \textit{when} the scalar filter is applied, namely, pre-filtering, post-filtering, and in-filtering \cite{DBLP:journals/corr/abs-2409-02571, DBLP:conf/icml/EngelsLYDS24, DBLP:journals/pacmmod/PatelKGZ24, DBLP:journals/corr/abs-2308-15014, DBLP:conf/www/GollapudiKSKBRL23, DBLP:journals/pacmmod/MohoneyPCMIMPR23}. However, as discussed in \textbf{C2} of \autoref{subsec:Motivation}, this classification is not enough to cover all algorithms, and too coarse to distinguish between algorithms.

In contrast, our framework focuses on the pruning behaviors of indices for hybrid queries. Since a hybrid query involves both vectors and scalars, it supports two types of pruning: \textit{vector pruning} and \textit{scalar pruning}. \textit{Vector pruning} refers to avoiding the computation of vector similarities for data points whose vectors are far from the query vector, typically achieved by searching on a vector index. \textit{Scalar pruning} refers to skipping vector similarity calculations for data points that do not satisfy the scalar filter, which can be efficiently implemented, as an example, using a bitmap generated by a scalar index to identify such points.

By emphasizing the dominant pruning behavior, our pruning-focused framework identifies four distinct strategies: \textit{vector-solely pruning} (VSP), \textit{vector-centric joint pruning} (VJP), \textit{scalar-centric joint pruning} (SJP), and \textit{scalar-solely pruning} (SSP). This finer-grained classification framework captures the core design principles of FANNS algorithms and overcomes the limitations of the existing framework by offering a more robust and extensible classification for both current and future FANNS algorithms, as detailed in the following subsections.

\subsection{VSP-based FANNS algorithms}
\label{subsec:VSP-based FANNS algorithms}

Vector-solely pruning (VSP) performs only vector pruning without any scalar pruning. The underlying rationale of VSP-based FANNS algorithms \cite{DBLP:conf/sigmod/YangLFW20, DBLP:conf/middleware/LiLGCNWC18, DBLP:journals/pvldb/WeiWWLZ0C20, DBLP:conf/sigmod/WangYGJXLWGLXYY21, DBLP:conf/osdi/ZhangXCSXCCH00Y23} (\textbf{A1}, \textbf{A2}) is that FANNS can be viewed as a direct extension of ANNS. These algorithms typically adapt existing vector indices with minimal modifications. VSP-based FANNS algorithms are suitable for scenarios with unselective scalar filters. However, due to the lack of scalar pruning, they may degrade to computing similarities for nearly all vectors when the scalar filter is highly selective.

\paragraph{A1: Post-Filtering Algorithm Family.} Post-Filtering Algorithm Family represents a class of methods \cite{DBLP:conf/sigmod/YangLFW20, DBLP:conf/middleware/LiLGCNWC18, DBLP:journals/pvldb/WeiWWLZ0C20, DBLP:conf/sigmod/WangYGJXLWGLXYY21} that directly use the top-k interface of a vector index designed for ANNS to retrieve the $K'$ nearest neighbors (\textit{$K'$-NN}), and then apply the scalar filter to find the final $K$ nearest neighbors (\textit{filtered $K$-NN}). For the choice of vector index, any type can be used, including IVF indices \cite{DBLP:conf/middleware/LiLGCNWC18, DBLP:conf/sigmod/YangLFW20, DBLP:journals/pvldb/WeiWWLZ0C20, DBLP:conf/sigmod/WangYGJXLWGLXYY21} that require minimal memory footprint, or graph indices \cite{DBLP:conf/sigmod/YangLFW20, DBLP:conf/sigmod/WangYGJXLWGLXYY21} that offer higher efficiency and accuracy. For the selection of $K'$, some methods choose a $K'$ much larger than $K$, hoping to retain at least $K$ elements after scalar filtering \cite{DBLP:journals/pvldb/WeiWWLZ0C20, DBLP:conf/sigmod/WangYGJXLWGLXYY21}; while others start with a $K'$ slightly larger than $K$ and iteratively increase it until at least $K$ data points are retained \cite{DBLP:conf/sigmod/YangLFW20}. Overall, Post-Filtering Algorithm Family is highly flexible and easy to implement since it can use any vector index off-the-shelf, but it faces the challenge of estimating the optimal $K'$ due to the unpredictable selectivity of the scalar filter during a specific search.

\paragraph{A2: VBase.} VBase \cite{DBLP:conf/osdi/ZhangXCSXCCH00Y23} optimizes the Post-Filtering Algorithm Family (\textbf{A1}) by dynamically selecting the optimal $K'$. Specifically, it modifies the result update process during the search over the ANNS index (\autoref{subsec:Background of ANNS Algorithms}) by adding only data points that satisfy the scalar filter to the result set. Besides, it introduces an improved termination check that requires the result set contain at least $K$ data points and meet the ``relaxed monotonicity'' condition \cite{DBLP:conf/osdi/ZhangXCSXCCH00Y23}, which ensures the result vectors are sufficiently similar to the query vector. With these modifications, the number of traversed data points during search serves as $K'$ in the Post-Filtering Algorithm Family, thereby addressing the difficulty in estimating $K'$, and this dynamically selected $K'$ has been proven to be optimal \cite{DBLP:conf/osdi/ZhangXCSXCCH00Y23}.

\subsection{VJP-based FANNS algorithms}
\label{subsec:VJP-based FANNS algorithms}

Vector-centric joint pruning (VJP) incorporates scalar pruning into the vector-pruning-centric search process. VJP-based FANNS algorithms \cite{DBLP:journals/corr/abs-2210-14958, DBLP:journals/pacmmod/PatelKGZ24, DBLP:conf/nips/WangLX0YN23, DBLP:conf/middleware/LiLGCNWC18, DBLP:journals/corr/abs-2401-08281, DBLP:conf/cikm/WuHQFLY22, DBLP:conf/www/GollapudiKSKBRL23, DBLP:journals/corr/abs-2308-15014, DBLP:journals/pacmmod/ZuoQZLD24, DBLP:journals/corr/abs-2409-02571} (\textbf{A3}–\textbf{A11}) further modify the search process over the ANNS index, so that the search direction is primarily guided by the similarity to the query vector, with adjustments made by the scalar filter, which may significantly reduce the number of vector similarity computations. However, the effectiveness of the scalar filter in assisting the guidance of search direction is not always guaranteed. Therefore, some studies go beyond modifying the search process by incorporating scalar information into the construction of indices \cite{DBLP:conf/nips/WangLX0YN23, DBLP:conf/cikm/WuHQFLY22, DBLP:conf/www/GollapudiKSKBRL23, DBLP:journals/pacmmod/ZuoQZLD24, DBLP:journals/corr/abs-2409-02571} (\textbf{A7}–\textbf{A11}). Overall, VJP-based FANNS algorithms tend to be more efficient than VSP-based FANNS algorithms (\autoref{subsec:VSP-based FANNS algorithms}) across varying levels of selectivity, but their results may be less reliable \cite{DBLP:journals/corr/abs-2210-14958, DBLP:journals/pacmmod/PatelKGZ24, DBLP:conf/nips/WangLX0YN23, DBLP:conf/cikm/WuHQFLY22} (\textbf{A3}, \textbf{A4}, \textbf{A7}, \textbf{A8}), and their applicability may be limited by their restrictive assumptions \cite{DBLP:conf/nips/WangLX0YN23, DBLP:conf/cikm/WuHQFLY22, DBLP:conf/www/GollapudiKSKBRL23, DBLP:journals/pacmmod/ZuoQZLD24, DBLP:journals/corr/abs-2409-02571} (\textbf{A7}–\textbf{A11}).

\paragraph{A3: AIRSHIP and A4: ACORN.} Attribute-Constrained Similarity Search on Proximity Graph (AIRSHIP) \cite{DBLP:journals/corr/abs-2210-14958} and ANN Constraint-Optimized Retrieval Network \newline (ACORN) \cite{DBLP:journals/pacmmod/PatelKGZ24} both incorporate scalar pruning into the search process over the graph index (\autoref{subsec:Background of ANNS Algorithms}). In AIRSHIP, data points that satisfy and do not satisfy the scalar filter are probabilistically visited during neighbor expansion, allowing exploiting satisfied data points to enhance efficiency, while exploring unsatisfied yet potentially useful data points for comprehensiveness. When updating the result set, it only adds data points that satisfy the scalar filter. In ACORN, only data points that satisfy the scalar filter are visited during neighbor expansion, aiming for thorough scalar pruning by traversing the \textit{predicate subgraph} \cite{DBLP:journals/pacmmod/PatelKGZ24}. While many graph indices \cite{DBLP:journals/pvldb/WangXY021} use a Relative Neighborhood Graph (RNG) \cite{DBLP:journals/pr/Toussaint80} approximation on Delaunay Graph (DG) \cite{DBLP:books/daglib/0031977} for sparsity, it is not suitable for ACORN since a sparse graph often leads to a disconnected \textit{predicate subgraph} \cite{DBLP:journals/pacmmod/PatelKGZ24}. Thus, ACORN retains the DG for a dense graph and designs a compression strategy to save memory. Overall, both algorithms are more efficient than VBase (\textbf{A2}) with graph indices, but uncertainty around the connectivity of the traversed subgraph makes search results potentially unreliable.

\paragraph{A5: Faiss-IVF and A6: CAPS.} Faiss-IVF \cite{DBLP:conf/middleware/LiLGCNWC18, DBLP:journals/corr/abs-2401-08281} and Constrained Approximate Partitioned Search (CAPS) \cite{DBLP:journals/corr/abs-2308-15014} both incorporate scalar pruning into the search process over the IVF index (\autoref{subsec:Background of ANNS Algorithms}). In Faiss-IVF, similarity calculations for vectors that do not satisfy the scalar filter are skipped during scanning. In CAPS, an \textit{attribute frequency tree} (AFT) \cite{DBLP:journals/corr/abs-2308-15014} is built for each cluster during index construction, with each AFT recursively partitioning the cluster based on its most frequently occurring scalar values. During search, the AFT narrows the scan scope within each cluster, then a scan similar to Faiss-IVF is performed in this refined scope. Overall, both algorithms are more efficient than VBase (\textbf{A2}) with IVF indices, and their memory footprint is smaller than that of graph indices, albeit with potentially lower search speed and accuracy.

\paragraph{A7: NHQ and A8: HQANN.} Native Hybrid Query \newline (NHQ) \cite{DBLP:conf/nips/WangLX0YN23} builds a composite graph index to enable joint pruning. It assumes discrete scalar values, and a \textit{simplified equality scalar filter} (\autoref{subsec:Definition of Hybrid Query}). For each data point, it transforms its scalar-tuple into a \textit{scalar vector} by encoding scalar values as numeric values, and then combines its vector with this \textit{scalar vector} to form a \textit{fusion vector}. Accordingly, it defines a \textit{fusion distance} metric to measure the similarity between \textit{fusion vectors}, such that data points with similar scalars and vectors have similar \textit{fusion vectors}. Consequently, FANNS over data points is transformed into ANNS over \textit{fusion vectors}, allowing both index construction and search accomplished using \textit{fusion vectors} and \textit{fusion distance}. Hybrid Query Approximate Nearest Neighbor Search (HQANN) \cite{DBLP:conf/cikm/WuHQFLY22} follows the core idea of NHQ, but introduces a different definition of the \textit{fusion distance}. Overall, both algorithms achieve high search efficiency by converting FANNS into ANNS under its assumptions, but uncertainty around the optimal form of \textit{fusion distance} makes search results potentially unreliable.

\paragraph{A9: Filtered-DiskANN.} Filtered-DiskANN \cite{DBLP:conf/www/GollapudiKSKBRL23} includes two similar methods, StitchedVamana and FilteredVamana, both incorporating scalar information into the construction and search process of the Vamana \cite{DBLP:conf/nips/SubramanyaDSKK19} graph index. It assumes discrete scalar values, and a \textit{simplified equality scalar filter} (\autoref{subsec:Definition of Hybrid Query}) whose conjunctive part in \autoref{eq:simplified scalar filter} only have one \textit{active sub-filter}. StitchedVamana constructs a separate graph for each scalar value over the subset of data points with that value in their scalar-tuple, then overlays these \textit{scalar-specific subgraphs} by unioning their edges and selectively retaining a limited number of edges to save memory. For a search using a scalar filter, it performs searches using each sub-filter, and then merges their results. For a search using a sub-filter, it only visits data points that satisfy the scalar filter during neighbor expansion, effectively performing ANNS on the \textit{scalar-specific subgraph}. FilteredVamana has a similar search process to StitchedVamana but constructs an approximate index by incrementally adding data points to an initially empty graph. For each newly added data point, it performs searches using each scalar value in the scalar-tuple of that data point as scalar filter, and the union of these results forms the candidate neighbors for connecting the new point. Overall, Filtered-DiskANN efficiently searches within subgraphs that meet the scalar filter like ACORN (\textbf{A4}), but achieves higher accuracy by explicitly constructing these subgraphs under its assumptions.

\paragraph{A10: SeRF.} Segment Graph for Range-Filtering (SeRF) \cite{DBLP:journals/pacmmod/ZuoQZLD24} follows the core idea of Filtered-DiskANN (\textbf{A9}) but operates under different assumptions. It assumes that each data point has only one scalar whose value is drawn from a discrete and orderable set, and a \textit{simplified range scalar filter} (\autoref{subsec:Definition of Hybrid Query}). For index construction, similar to Filtered-DiskANN which builds a graph by approximately overlaying \textit{scalar-specific subgraphs} for all possible scalar values, SeRF constructs a graph by approximately overlaying \textit{range-specific subgraphs} for all possible scalar value ranges (in the form of $[a,b], a \le b$). Assuming there are $n$ distinct scalar values (the same as the number of data points), Filtered-DiskANN overlays $n$ subgraphs and achieves a worst-case space complexity of $O(Mn)$ by limiting each point to $M$ neighbors \cite{DBLP:conf/www/GollapudiKSKBRL23}, while SeRF overlays $n^{2}$ subgraphs and has a worst-case space complexity of $O(M n^{2})$ due to its ``lossless compression'' \cite{DBLP:journals/pacmmod/ZuoQZLD24}. For FANNS, similar to Filtered-DiskANN which effectively performs searches on \textit{scalar-specific subgraphs}, SeRF effectively performs searches on \textit{range-specific subgraphs}. Overall, SeRF also provides higher efficiency and accuracy than ACORN (\textbf{A4}) under its assumptions, but it suffers from a high memory footprint.

\paragraph{A11: iRangeGraph.} Improvising Range-dedicated Graphs (iRangeGraph) \cite{DBLP:journals/corr/abs-2409-02571} follows the same assumptions and search strategy as SeRF (\textbf{A10}), but adopts a different approach for index construction. Rather than overlaying numerous \textit{range-specific subgraphs}, iRangeGraph constructs a moderate number of \textit{range-specific subgraphs} and organizes them in a \textit{segment tree} structure \cite{DBLP:journals/corr/abs-2409-02571}. In this \textit{segment tree}, each node represents a range and stores the range-specific subgraph constructed over the subset of data points with scalar values in that range, which results in a worst-case space complexity of $O(Mn\log{n})$. During search, as each data point appears in \textit{range-specific subgraphs} at multiple levels of the tree, its neighbors for expansion are the union of its neighbors across all the tree levels. Overall, iRangeGraph achieves a smaller memory footprint than SeRF (\textbf{A10}) while ensuring search efficiency and accuracy under its assumptions.

\subsection{SSP-based FANNS algorithms}
\label{subsec:SSP-based FANNS algorithms}

Scalar-Solely Pruning (SSP) performs only scalar pruning without any vector pruning. SSP-based FANNS algorithms \cite{DBLP:journals/pvldb/WeiWWLZ0C20, DBLP:conf/sigmod/WangYGJXLWGLXYY21, DBLP:conf/osdi/ZhangXCSXCCH00Y23, DBLP:journals/pacmmod/PatelKGZ24} (\textbf{A12}) are easy to implement and memory efficient without the need for vector indices, and are suitable for scenarios with selective scalar filters. However, due to the lack of vector pruning, they may degrade to computing similarities for nearly all vectors when the scalar filter is unselective.

\paragraph{A12: Pre-Filtering Algorithm Family.} Pre-Filtering Algorithm Family represents a class of methods \cite{DBLP:journals/pvldb/WeiWWLZ0C20, DBLP:conf/sigmod/WangYGJXLWGLXYY21, DBLP:conf/osdi/ZhangXCSXCCH00Y23, DBLP:journals/pacmmod/PatelKGZ24} that first apply the scalar filter to retrieve a subset of data points (\textit{filtered subset}), and then find the \textit{filtered $K$-NN} by calculating similarities for all vectors in this subset through brute-force scan. Since the number of possible \textit{filtered subsets} can be extremely large or even uncountable (e.g., when the scalar filter is based on regular expressions), it is impractical to construct vector indices for all possible \textit{filtered subsets} to accelerate the search process, making vector pruning infeasible. However, the efficiency of brute-force scan can be improved. For example, AnalyticDB-V (ADBV) \cite{DBLP:journals/pvldb/WeiWWLZ0C20} trades accuracy for efficiency by pre-compressing data vectors using Voronoi graph product quantization and performing asymmetric distance computation, in which query vectors remain uncompressed while data vectors are compressed.

\subsection{SJP-based FANNS algorithms}
\label{subsec:SJP-based FANNS algorithms}

Scalar-Centric Joint Pruning (SJP) incorporates vector pruning into the scalar-pruning-centric search process. To achieve this, SJP-based FANNS algorithms \cite{DBLP:conf/sigmod/WangYGJXLWGLXYY21, DBLP:journals/pacmmod/MohoneyPCMIMPR23, DBLP:journals/concurrency/XuLWX20, ACM:10.1145/3698822, DBLP:conf/icml/EngelsLYDS24} (\textbf{A13}–\textbf{A17}) define rules to select a limited number of \textit{filtered subsets}, and then build vector indices for each selected subset. During search, they first use part of the scalar filter to coarsely retrieve some of these preselected subsets (\textit{partially filtered subsets}), and then carry out hybrid searches on them to obtain the final \textit{filtered $K$-NN}. Notably, the \textit{partially filtered subsets} may contain data points that do not satisfy the scalar filter, so scalar pruning is not as complete as in SSP-based FANNS algorithms (\autoref{subsec:SSP-based FANNS algorithms}), but since vector indices are built on these subsets, vector pruning can be leveraged to accelerate the search process. Overall, SJP-based FANNS algorithms require careful selection of subsets to build vector indices, and their applicability is all limited by their restrictive assumptions.

\paragraph{A13: Milvus-Partition and A14: HQI.} Milvus-Partition \cite{DBLP:conf/sigmod/WangYGJXLWGLXYY21} and Hybrid Query Index (HQI) \cite{DBLP:journals/pacmmod/MohoneyPCMIMPR23} both partition the dataset into disjoint subsets based on workload characteristics and build vector indices for each subset. During search, they both retrieve \textit{partially filtered subsets} and apply one of the VSP-, VJP-, or SSP-based FANNS algorithms for each subset, and finally merge the results to obtain the final \textit{filtered $K$-NN}. However, the criteria and structure of dataset partitioning differ between them. In Milvus-Partition, partitioning is scalar-based and single-layered. It first identifies the most frequently used scalar from prior workload, and then evenly divides the dataset into subsets based on the values of this scalar. In HQI, partitioning is filter-based and multi-layered. It first identifies several frequently used scalar filters from prior workload, and then constructs an \textit{extended qd-tree} \cite{DBLP:journals/pacmmod/MohoneyPCMIMPR23} by iteratively partitioning the dataset according to these filters. Overall, both algorithms achieve high search efficiency when workload characteristics are stable, but their applicability is limited by the need for the prior workload information to guide the partitioning process.

\paragraph{A15: MA-NSW.} Multiattribute ANNS based on Navigable Small World (MA-NSW) \cite{DBLP:journals/concurrency/XuLWX20} constructs multiple NSW \cite{DBLP:journals/is/MalkovPLK14} graph indices based on the values of the scalar-tuples. It assumes discrete scalar values, and a \textit{simplified equality scalar filter} (\autoref{subsec:Definition of Hybrid Query}). For index construction, MA-NSW first defines a containment relationship between two scalar-tuples $\mathbf{s}_{1}$ and $\mathbf{s}_{2}$, where $\mathbf{s}_{1}$ is included by $\mathbf{s}_{2}$ ($\mathbf{s}_{1} \subseteq \mathbf{s}_{2}$) if $\mathbf{s}_{2}$ has at least one scalar with a NULL value and all other scalar values are identical to those of $\mathbf{s}_{1}$. Assuming the scalar schema has $m$ scalars, each with $m_{i}$ distinct values including the NULL value, there will be $\Pi_{i=1}^{m} m_{i}$ possible distinct scalar-tuples. For each observed scalar-tuple, MA-NSW identifies a subset of data points whose scalar-tuples are either identical to or included by the given scalar-tuple, and then constructs an NSW on this subset. Since a single NSW has a space complexity of $O(Mn)$, the worst-case space complexity of MA-NSW is $O(Mn^{m+1})$. For a search using a scalar filter, where each conjunctive sub-filter corresponds to a subset with a prebuilt NSW, MA-NSW retrieves all relevant subsets, performs ANNS on each NSW, and finally merges the results. Overall, despite the high efficiency of MA-NSW under its assumptions, its memory footprint can be prohibitively large.

\paragraph{A16: UNG.} Unified Navigating Graph (UNG) \cite{ACM:10.1145/3698822} constructs a single graph index based on the values of the scalar-tuples. It has the same assumptions and containment relationship definition as in MA-NSW (\textbf{A15}). Additionally, it defines a minimal containment relationship between two scalars $\mathbf{s}_{1}$ and $\mathbf{s}_{2}$, where $\mathbf{s}_{1}$ is minimally included by $\mathbf{s}_{2}$ if $\mathbf{s}_{1} \subseteq \mathbf{s}_{2}$ and no other scalar $\mathbf{s}_{3}$ exists such that $\mathbf{s}_{1} \subseteq \mathbf{s}_{3} \subseteq \mathbf{s}_{2}$. During index construction, for each scalar-tuple, UNG identifies the subset of data points whose scalar-tuples are identical to the given scalar-tuple, and then constructs a graph index on this subset. After that, if the scalar-tuple of one subset is minimally included by that of another, UNG selectively adds directed edges from some data points in the latter subset to some in the former subset, organizing these subsets in a manner similar to a \textit{prefix tree} structure \cite{ACM:10.1145/3698822}. Since all the subsets are disjoint from each other and the number of added edges is limited, the worst-case space complexity of UNG is $O(Mn)$. For a search using a scalar filter, where each conjunctive sub-filter corresponds to a group of linked subsets with a \textit{concatenated graph index} by combining graph indices of these subsets with directed edges connecting these subsets, UNG retrieves all relevant groups, performs ANNS on each \textit{concatenated graph index}, and finally merges the results. Overall, UNG achieves a small memory footprint and high efficiency when its assumptions are satisfied and containment relationships are abundant.

\paragraph{A17: WST.} Window Search Tree (WST) \cite{DBLP:conf/icml/EngelsLYDS24} shares the same assumptions and a similar index structure with iRangeGraph (\textbf{A11}), but introduces four different search methods: VamanaWST, OptimizedPostfiltering, ThreeSplit, and SuperPostfiltering. WST organizes \textit{range-specific subgraphs} in an extended form of \textit{segment tree} \cite{DBLP:conf/icml/EngelsLYDS24}, where each node represents a range and its children recursively divide this range into $\beta$ sub-ranges, which reduces to the standard \textit{segment tree} when $\beta = 2$. VamanaWST recursively searches the WST from top to bottom to identify a set of nodes whose ranges are disjoint and the union of their ranges is the query range (i.e., the scalar filter), then performs ANNS on each node, and finally merges the results. OptimizedPostfiltering selects a single node with the smallest range that contains the query range and applies one of the VSP-, VJP-, or SSP-based algorithms on this node to obtain the result. ThreeSplit first selects a node with the largest range contained within the query range and performs ANNS on this node, then applies OptimizedPostfiltering to the remaining portions of the query range, and finally merges the results. SuperPostfiltering is similar to OptimizedPostfiltering but does not rely on \textit{range-specific subgraphs} in the WST; instead, it constructs an arbitrary set of \textit{range-specific subgraphs} to achieve an expected ``small blowup'' \cite{DBLP:conf/icml/EngelsLYDS24}. Overall, all four methods achieve small memory footprint and high search efficiency under given assumptions.

\section{Review of Hybrid Datasets}
\label{sec:Review of Hybrid Datasets}

\begin{table*}[t]
  \caption{Summary of Selected Hybrid Datasets}
  \label{tab:dataset}
  \centering
  \renewcommand{\arraystretch}{1.3}
  \resizebox{\textwidth}{!}{
  \begin{threeparttable}
    \begin{tabular}{l|c|c|c|c|l}
    \hline
    \multicolumn{1}{c|}{\multirow{2}{*}{\textbf{Dataset}}} & \multicolumn{1}{c|}{\multirow{2}{*}{\textbf{\#Points}}} & \multicolumn{2}{c|}{\textbf{Vector}} & \multicolumn{1}{c|}{\textbf{Scalar-Tuple}} & \multicolumn{1}{c}{\multirow{2}{*}{\textbf{Used in}}} \\
    \cline{3-5}
    & & \multicolumn{1}{c|}{\textbf{Source}} & \multicolumn{1}{c|}{\textbf{\#Dimensions}} & \multicolumn{1}{c|}{\textbf{\#Organic}} & \\
    \hline
    \noalign{\vskip 2pt}
    \hline

    \textbf{D1: }SIFT-1M & 1,000,000 & Image & 128 & 0 & \cite{DBLP:journals/concurrency/XuLWX20, DBLP:conf/sigmod/YangLFW20, DBLP:journals/pvldb/WeiWWLZ0C20, DBLP:conf/sigmod/WangYGJXLWGLXYY21, DBLP:journals/corr/abs-2210-14958, DBLP:conf/nips/WangLX0YN23, DBLP:conf/cikm/WuHQFLY22, DBLP:conf/www/GollapudiKSKBRL23, DBLP:journals/pacmmod/MohoneyPCMIMPR23, DBLP:journals/corr/abs-2308-15014, DBLP:journals/pacmmod/PatelKGZ24, DBLP:conf/icml/EngelsLYDS24, ACM:10.1145/3698822} \\ \hline
    \textbf{D2: }GIST-1M & 1,000,000 & Image & 960 & 0 & \cite{DBLP:journals/concurrency/XuLWX20, DBLP:conf/sigmod/YangLFW20,  DBLP:conf/nips/WangLX0YN23, DBLP:conf/cikm/WuHQFLY22, DBLP:journals/pacmmod/MohoneyPCMIMPR23, DBLP:conf/icml/EngelsLYDS24, ACM:10.1145/3698822} \\ \hline
    \textbf{D3: }Deep-10M & 10,000,000 & Image & 96 & 0 & \cite{DBLP:journals/pvldb/WeiWWLZ0C20, DBLP:conf/sigmod/WangYGJXLWGLXYY21, DBLP:conf/cikm/WuHQFLY22, DBLP:journals/pacmmod/ZuoQZLD24, DBLP:conf/icml/EngelsLYDS24} \\ \hline
    \textbf{D4: }MNIST-8M & 8,100,000 & Image & 784 & 1 & \cite{DBLP:journals/concurrency/XuLWX20, DBLP:journals/corr/abs-2210-14958} \\ \hline
    \textbf{D5: }MTG & 40,274 & Image & 1,152 & 21 & \cite{ACM:10.1145/3698822} \\ \hline
    \textbf{D6: }GloVe-Twitter & 1,183,514 & Word & 25; 50; 100; 200 & 1 & \cite{DBLP:journals/concurrency/XuLWX20, DBLP:conf/nips/WangLX0YN23, DBLP:conf/cikm/WuHQFLY22, DBLP:journals/corr/abs-2308-15014, DBLP:conf/icml/EngelsLYDS24, ACM:10.1145/3698822} \\ \hline
    \textbf{D7: }GloVe-Crawl & 1,989,995 & Word & 300 & 1 & \cite{DBLP:conf/nips/WangLX0YN23, DBLP:journals/corr/abs-2308-15014, ACM:10.1145/3698822} \\ \hline
    \multicolumn{1}{l|}{\multirow{2}{*}{\textbf{D8: }LAION-1M}} & \multicolumn{1}{c|}{\multirow{2}{*}{1,000,448}} & \multicolumn{1}{c|}{Image} & \multicolumn{1}{c|}{512} & \multicolumn{1}{c|}{\multirow{2}{*}{15}} & \multicolumn{1}{l}{\multirow{2}{*}{\cite{DBLP:journals/pacmmod/ZuoQZLD24, DBLP:journals/corr/abs-2409-02571}}} \\ \cline{3-4}
    & & \multicolumn{1}{c|}{Text} & \multicolumn{1}{c|}{512} & & \\ \hline
    \multicolumn{1}{l|}{\multirow{2}{*}{\textbf{D9: }YouTube}} & \multicolumn{1}{c|}{\multirow{2}{*}{6,134,598}} & \multicolumn{1}{c|}{Video} & \multicolumn{1}{c|}{1,024} & \multicolumn{1}{c|}{\multirow{2}{*}{3}} & \multicolumn{1}{l}{\multirow{2}{*}{\cite{DBLP:journals/pacmmod/PatelKGZ24, ACM:10.1145/3698822}}} \\ \cline{3-4}
    & & \multicolumn{1}{c|}{Audio} & \multicolumn{1}{c|}{128} & & \\ \hline
    \end{tabular}
  \end{threeparttable}
  }
\end{table*}

In this section, we discuss the construction strategies of existing hybrid datasets, and present several examples to detail their contents, as summarized in \autoref{tab:dataset}.

\subsection{Hybrid Dataset Construction}
\label{subsec:Hybrid Dataset Construction}

Hybrid datasets can be seen as vector datasets with added scalars. However, unlike the evaluation of ANNS, which benefits from a standardized set of vector datasets \cite{DBLP:journals/pami/JegouDS11, DBLP:conf/icassp/JegouTDA11, DBLP:conf/cvpr/YandexL16, DBLP:conf/emnlp/PenningtonSM14} provided by standard benchmarks like ANN-Benchmarks \cite{DBLP:journals/is/AumullerBF20}, the evaluation of FANNS lacks a comparable standard for hybrid datasets. As a result, existing FANNS studies often customize their own hybrid datasets. 

Some studies synthesize scalars such as generating random values from a uniform distribution \cite{DBLP:journals/concurrency/XuLWX20, DBLP:journals/pvldb/WeiWWLZ0C20, DBLP:conf/sigmod/YangLFW20, DBLP:conf/sigmod/WangYGJXLWGLXYY21, DBLP:journals/corr/abs-2210-14958, DBLP:conf/nips/WangLX0YN23, DBLP:conf/cikm/WuHQFLY22, DBLP:conf/www/GollapudiKSKBRL23, DBLP:journals/pacmmod/MohoneyPCMIMPR23, DBLP:journals/corr/abs-2308-15014, DBLP:journals/pacmmod/PatelKGZ24, DBLP:journals/pacmmod/ZuoQZLD24, DBLP:conf/icml/EngelsLYDS24, ACM:10.1145/3698822}, while others use ``organic'' scalars that already exists in the original data sources \cite{DBLP:journals/pvldb/WeiWWLZ0C20, DBLP:journals/corr/abs-2210-14958, DBLP:conf/osdi/ZhangXCSXCCH00Y23, DBLP:conf/www/GollapudiKSKBRL23, DBLP:journals/pacmmod/PatelKGZ24, DBLP:journals/pacmmod/ZuoQZLD24, DBLP:conf/icml/EngelsLYDS24, DBLP:journals/corr/abs-2409-02571}. In the former case, the vectors can be sourced from any vector dataset, including those in the ANN-Benchmarks. In the latter case, the original data source is typically collected by crawling web pages, where unstructured data (e.g., images, texts, or audio) are used to extract the feature vectors through a pre-trained model (e.g., CNN \cite{DBLP:journals/pieee/LeCunBBH98, DBLP:journals/corr/SimonyanZ14a, DBLP:conf/cvpr/HeZRS16}, Transformer \cite{DBLP:conf/nips/VaswaniSPUJGKP17, DBLP:conf/nips/BrownMRSKDNSSAA20, DBLP:conf/naacl/DevlinCLT19}, or VGGish \footnote{\url{https://github.com/tensorflow/models/tree/master/research/audioset/vggish}}), and structured data (e.g., publication dates, keywords, and likes) serve as the corresponding scalars.

Based on the above analysis, we categorize existing hybrid datasets into two types: (1) \textit{synthesized hybrid datasets}, which contain only synthesized scalars; and (2) \textit{organic hybrid datasets}, which contain organic scalars, either exclusively or in combination with synthesized ones.

\subsection{Representative Hybrid Datasets}
\label{subsec:Representative Hybrid Datasets}

Among existing hybrid datasets, we select nine representative examples (\textbf{D1}-\textbf{D9}) for detailed description, including three \textit{synthesized hybrid datasets} (\textbf{D1}-\textbf{D3}) whose vectors are sourced from ANN-Benchmarks, and six \textit{organic hybrid datasets} (\textbf{D4}-\textbf{D9}) with traceable raw data sources. A concise summary of these datasets is provided in \autoref{tab:dataset}, including information on data size, vector source and dimensionality, number of organic scalars, and their usage in recent studies.

\paragraph{D1: SIFT-1M, D2: GIST-1M, and D3: Deep-10M.} They are widely-used vector datasets from ANN-Benchmarks. SIFT-1M \footnote{\url{http://corpus-texmex.irisa.fr/} \label{footnote:bigann}} \cite{DBLP:journals/pami/JegouDS11} contains 1 million 128-dimensional vectors, extracted from the INRIA Holidays images \cite{DBLP:conf/eccv/JegouDS08} using local SIFT descriptors \cite{DBLP:journals/ijcv/Lowe04}. GIST-1M \textsuperscript{\ref{footnote:bigann}} \cite{DBLP:journals/pami/JegouDS11} contains 1 million 960-dimensional vectors, extracted from the INRIA Holidays images \cite{DBLP:conf/eccv/JegouDS08} using global GIST descriptors \cite{DBLP:journals/ijcv/OlivaT01}. Deep-10M \footnote{\url{https://research.yandex.com/blog/benchmarks-for-billion-scale-similarity-search} \label{footnote:deep}} \cite{DBLP:conf/cvpr/YandexL16} contains 10 million 96-dimensional vectors, extracted from the images on the Web using the GoogLeNet model \cite{DBLP:journals/corr/SzegedyLJSRAEVR14}. Original SIFT, GIST and Deep do not have organic scalars, so synthesized scalars are generated to make them hybrid datasets. Notably, SIFT and Deep also have 1 billion versions, namely SIFT-1B \textsuperscript{\ref{footnote:bigann}} and DEEP-1B \textsuperscript{\ref{footnote:deep}}.

\paragraph{D4: MNIST-8M.} MNIST-8M \footnote{\url{https://www.csie.ntu.edu.tw/~cjlin/libsvmtools/datasets/multiclass.html\#mnist8m}} \cite{loosli-canu-bottou-2006} contains 8,100,000 data points, each comprising a 784-dimensional vector representation of a handwritten digit image generated using SVMs \cite{DBLP:journals/ml/CortesV95}, along with an integer label between 0 and 9. These images are derived from the infinite MNIST dataset (InfiMNIST) \footnote{\url{https://leon.bottou.org/projects/infimnist}}, which extends the original MNIST dataset (MNIST) \cite{DBLP:journals/spm/Deng12} by dynamically generating new samples through careful elastic deformations, theoretically enabling the creation of an unlimited number of images. Notably, from InfiMNIST, 10,000 samples ranging from 0 to 9,999 form the MNIST testing set, 60,000 samples ranging from 10,000 to 69,999 form the MNIST training set, and 8,100,000 samples ranging from 10,000 to 8,109,999 form the MNIST-8M.

\paragraph{D5: MTG.} MTG \footnote{\url{https://huggingface.co/datasets/TrevorJS/mtg-scryfall-cropped-art-embeddings-open-clip-ViT-SO400M-14-SigLIP-384}} contains 40,274 data points, each comprising a 1,152-dimensional vector representation of a ``Magic: The Gathering'' \footnote{\url{https://scryfall.com/}} card image generated using the OpenCLIP model \cite{DBLP:conf/cvpr/ChertiBWWIGSSJ23}, along with 21 scalars, such as ``artist'', ``rarity'', and ``toughness''. Notably, the content of ``image'' scalar is the raw card image, and several scalars ending with ``uri'' provide traceable links to the websites from which the dataset was crawled.

\paragraph{D6: GloVe-Twitter and D7: Glove-Crawl.}  GloVe-Twitter \footnote{\url{https://nlp.stanford.edu/projects/glove/} \label{footnote:glove}} and GloVe-Crawl \textsuperscript{\ref{footnote:glove}} both contain word vectors generated using the GloVe algorithm \cite{DBLP:conf/emnlp/PenningtonSM14} from the Twitter corpus and the Common Crawl corpus, respectively. GloVe-Twitter contains 1,183,514 unique words, with corresponding word vectors generated in four different dimensions: 25, 50, 100, and 200. This means that GloVe-Twitter provides four sets of word vectors, each containing 1,183,514 vectors for the respective dimensions. In contrast, GloVe-Crawl contains 1,989,995 unique words, with word vectors generated in a single fixed dimension of 300.


\paragraph{D8: LAION-1M.} LAION-1M \footnote{\url{https://deploy.laion.ai/}} contains the first 1,000,448 data points from LAION-400M \cite{DBLP:journals/corr/abs-2111-02114}, with each data point comprising 2 vectors and 15 scalars. Each vector pair includes a 512-dimensional image embedding and a corresponding 512-dimensional text embedding, both generated from the Common Crawl corpus using the same CLIP model \cite{DBLP:conf/icml/RadfordKHRGASAM21}. The scalar part includes the image URL and associated metadata, such as the image's width and height. Notably, LAION-5B \cite{DBLP:conf/nips/SchuhmannBVGWCC22} is also available, containing a significantly larger dataset of 5,526,641,167 data points, while maintaining a structure similar to the previous versions.

\paragraph{D9: YouTube.} YouTube \footnote{\url{https://research.google.com/youtube8m/download.html}} \cite{DBLP:conf/nips/SchuhmannBVGWCC22}, an updated version of YouTube-8M \cite{DBLP:journals/corr/Abu-El-HaijaKLN16}, contains 6,134,598 data points, each comprising 2 vectors and 3 scalars. Each vector pair includes a 1,024-dimensional video embedding and a corresponding 128-dimensional audio embedding, generated from the YouTube corpus using Inception model \cite{DBLP:conf/icml/IoffeS15} and VGGish model \footnote{\url{https://github.com/tensorflow/models/blob/master/research/audioset/vggish/README.md}}, respectively. The scalars include the sample ID, video URL and video labels. 

\section{The Distribution Factor for Query Difficulty}
\label{sec:The Distribution Factor for Query Difficulty}

Understanding query difficulty is essential for analyzing algorithm performance and designing evaluation benchmarks. Query difficulty influences both the efficiency and accuracy of an algorithm. Specifically, more difficult queries tend to increase search time and decrease \textit{recall}. Multiple factors may contribute to query difficulty. Identifying these factors not only helps explain performance fluctuations across different queries from various perspectives, but also enables the construction of evaluation benchmarks across a wider range of difficulty levels through their combination.

This section goes beyond the \textit{selectivity} factor to explore the role of the \textit{distribution} factor in the difficulty of hybrid queries. We begin by motivating the incorporation of the \textit{distribution} factor by examining real-world hybrid datasets. Next, we conduct carefully designed experiments to assess the impact of the \textit{distribution} factor on query difficulty, followed by both qualitative visualizations and quantitative measurements to explain the results. Finally, we propose a schema towards more comprehensive FANNS algorithm evaluation by incorporating both \textit{selectivity} and \textit{distribution} factors.
\begin{figure*}
  \centering
  \includegraphics[width=0.93\linewidth]{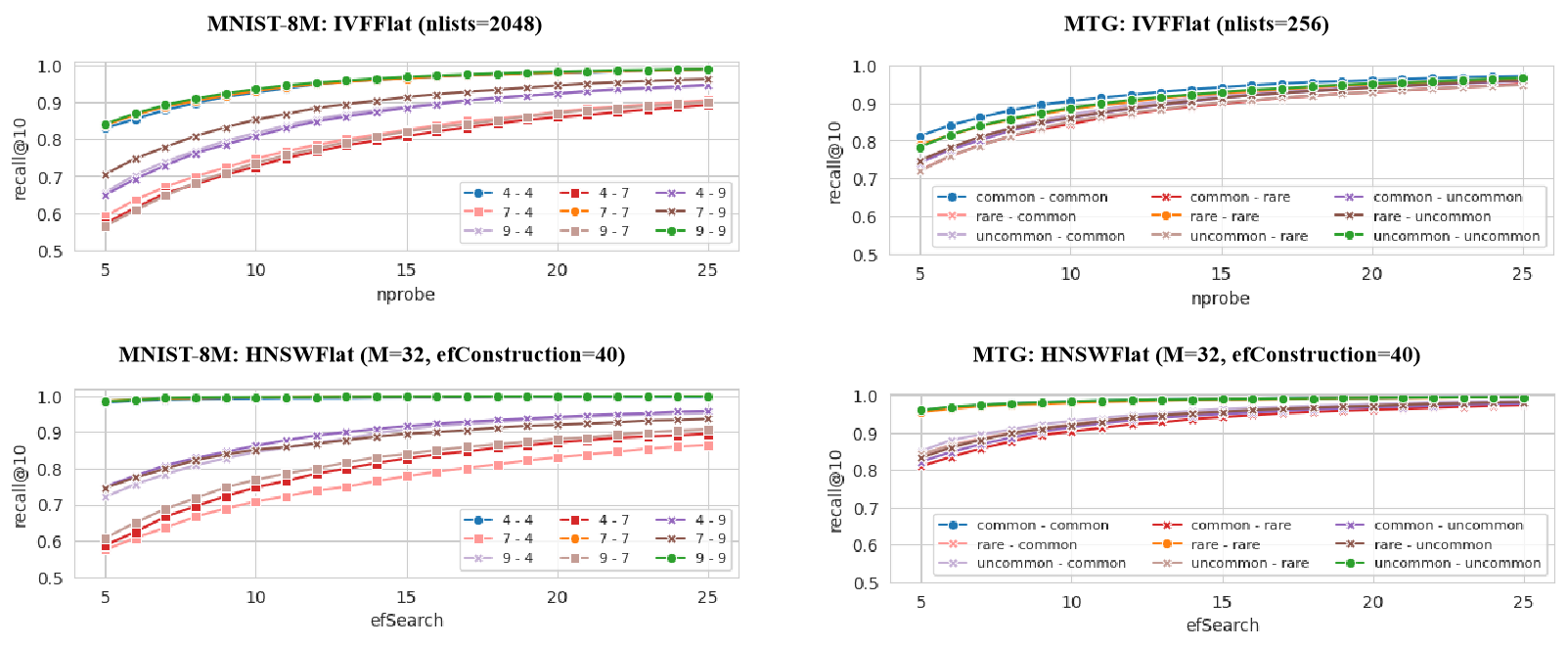}
  \caption{Performance evaluations on hybrid queries with oracle partition indices. Each ``x-y'' is a set of hybrid queries, where the scalar value of base vectors is x and the scalar value of query vectors is y. Each set of base vectors is indexed using IVF and HNSW. The distribution relationship of each hybrid query set is ID, POD, or OOD, represented by circle, cross, or square.}
  \label{fig:oracle}
\end{figure*}

\begin{figure*}
  \centering
  \includegraphics[width=0.93\linewidth]{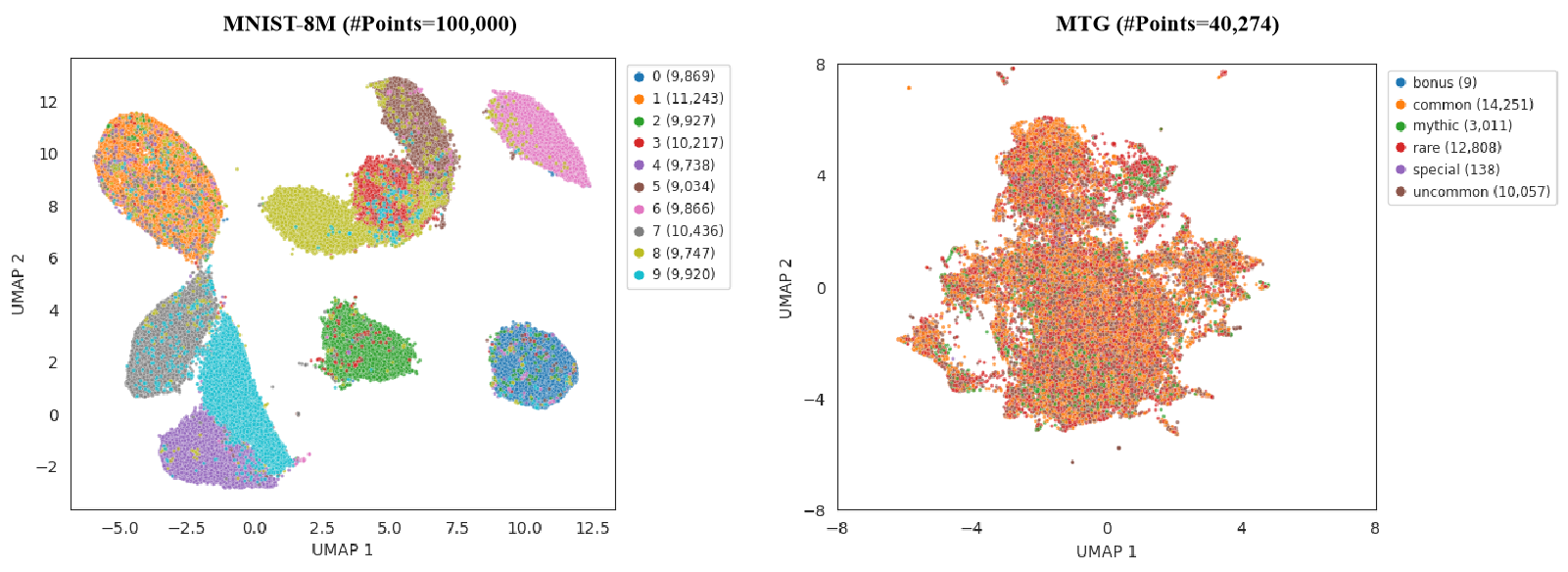}
  \caption{UMAP visualizations of MNIST-8M (Left) and MTG (Right). Each \textit{filtered subset} has a 2-dimensional distribution. In MNIST-8M, the distributions of each \textit{filtered subset} are well-separated, exhibiting clustering behavior, while in MTG, the distributions of all \textit{filtered subsets} are highly overlapping, sharing a similar overall distribution.}
  \label{fig:umap}
\end{figure*}

\begin{figure*}
  \centering
  \includegraphics[width=0.93\linewidth]{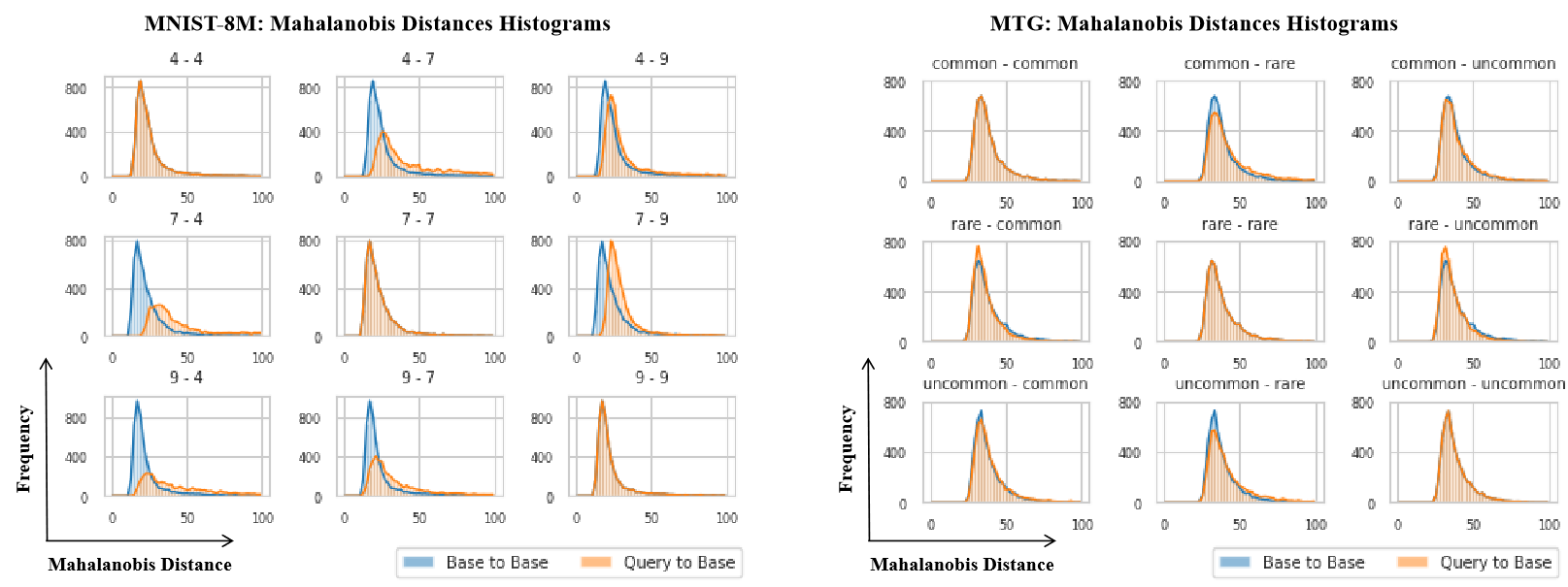}
  \caption{Mahalanobis Distance Histograms of MNIST-8M (Left) and MTG (Right). Each subgraph titled ``x-y'' shows the histograms of both ``x-y'' (in orange) and ``x-x'' (in blue), illustrating the distribution shift between query vectors and base vectors when sampled from different \textit{filtered subsets}, compared to when they are sampled from the same \textit{filtered subset}.}
  \label{fig:mahalanobis}
\end{figure*}

\subsection{Incorporation of Distribution Factor: Motivation}
\label{subsec:Incorporation of Distribution Factor: Motivation}

The \textit{selectivity} measures the proportion of data points excluded by the scalar filter (\autoref{subsec:Definitions of Evaluation Metrics}). It is currently the only factor used to evaluate the difficulty of hybrid queries. As discussed in \autoref{sec:Review of FANNS Algorithms}, hybrid queries with high \textit{selectivity} are more difficult for VSP-based FANNS algorithms, whereas those with low \textit{selectivity} are more difficult for SSP-based algorithms. In the meantime, VJP-based and SJP-based algorithms are robust across varying levels of \textit{selectivity} if their corresponding assumptions are satisfied.

In real-world scenarios, hybrid datasets often exhibit clustering behavior, where data points with similar scalars tend to form distinct clusters in the vector space. For instance, in e-commerce platforms, products from the same brand or sharing a common style often exhibit clustered embeddings, reflecting their inherent similarity \cite{DBLP:conf/sigir/McAuleyTSH15}. In the domain of news articles, temporal proximity can lead to clustering, as articles covering the same event within a short timeframe tend to have highly similar content and embeddings \cite{DBLP:journals/air/RazaD22}. 

This motivates us to incorporate the \textit{distribution} factor into the hybrid query difficulty. In the high dimensional vector space, each group of vectors has a high dimensional distribution. The \textit{distribution} factor refers to the relationship between distributions of two sets of vectors. If query vectors and base vectors are sampled from two different \textit{filtered subsets} within a clustered dataset, the relationship between their distributions is likely to be Out-of-Distribution (OOD) \cite{DBLP:journals/corr/abs-2211-12850, DBLP:journals/pvldb/ChenZHJW24}, leading to difficult hybrid queries, because query vectors are far from their nearest neighbors in the base vectors, and the nearest neighbors are also distant from each another \cite{DBLP:journals/pvldb/ChenZHJW24}. Conversely, if query vectors and base vectors are sampled from the same \textit{filtered subset} within a clustered dataset, or sampled from a dataset without clustering behavior, their distribution relationship is likely to be In-Distribution (ID) \cite{DBLP:journals/corr/abs-2211-12850, DBLP:journals/pvldb/ChenZHJW24}, resulting in easier hybrid queries.

\subsection{Impact of Distribution Factor: Experiments and Explanations}
\label{subsec:Impact of Distribution Factor: Experiments and Explanations}

To illustrate how the \textit{distribution} factor impacts the difficulty of hybrid queries, we conduct experiments on two hybrid datasets introduced in \autoref{sec:Review of Hybrid Datasets}: MNIST-8M, which exhibits clustering behavior, and MTG, which does not. For demonstration purposes, we consider only one vector column and one scalar column in each dataset, with the scalar filter checking whether the scalar equals a specific value. Concretely, for MNIST-8M, we use the first 1,000,000 data points and select \texttt{digit} as the scalar attribute. For MTG, we use all 40,274 data points and choose \texttt{rarity} as the scalar attribute. Under this setup, each \textit{filtered subset} consists of vectors associated with a particular scalar value, and different \textit{filtered subsets} exhibit distinct distributions. To evaluate whether two \textit{filtered subsets} are OOD, ID, or in an intermediate state (referred to as Partially-Overlapping-Distribution, POD), we design hybrid queries across \textit{filtered subsets}, identifying their relationships based on the performance gap relative to a baseline.

\paragraph{Experiment Design.} Suppose the chosen scalar $\mathbb{S}_{c}$ has $|\mathbb{S}_{c}|$ unique values. Let ``$s_{base} - s_{query}$'' be a set of hybrid queries, where base vectors are sampled from the \textit{filtered subset} where the scalar value is $s_{base} \in \mathbb{S}_{c}$, and query vectors are sampled from the \textit{filtered subset} where the scalar value is $s_{query} \in \mathbb{S}_{c}$. This results in a total of $|\mathbb{S}_{c}|^2$ hybrid query sets. Among these, hybrid queries where $s_{base} = s_{query}$ are referred to as \textit{baseline hybrid queries}, with $|\mathbb{S}_{c}|$ such queries in total. To achieve theoretically optimal search performance for each hybrid query, an ANN index must exist for the corresponding base vectors. This index, referred to as the \textit{oracle partition index} \cite{DBLP:journals/pacmmod/PatelKGZ24}, transforms FANNS over the entire dataset into ANNS within base vectors.

For demonstration purposes, we select 3 scalar values from each dataset: $\mathbb{S}_{c}=\{\text{\texttt{4}}, \text{\texttt{7}}, \text{\texttt{9}}\}$ from MNIST-8M, and $\mathbb{S}_{c}=\{\text{\texttt{common}}, \text{\texttt{rare}}, \text{\texttt{uncommon}}\}$ from MTG. Each hybrid query set contains 1,000 randomly sampled vectors from each \textit{filtered subset}, with $k=10$. Each set of base vectors has 2 types of ANN indices—IVFFlat and HNSWFlat—constructed using the Faiss library \footnote{\url{https://github.com/facebookresearch/faiss/tree/v1.9.0}}. For IVFFlat, the construction parameter \texttt{nlists} (the number of inverted lists) is set to 2,048 and 256, respectively, as recommended by the library. The search parameter \texttt{nprobe} (the number of inverted lists visited during a query) is varied from 5 to 25, respectively, and the average \texttt{recall@10} is measured for each hybrid query set. For HNSWFlat, the construction parameters \texttt{M} (the number of neighbors in the graph) and \texttt{efConstruction} (the depth of exploration during construction) are both set to 32 and 40, following the library’s default settings. The search parameter \texttt{efSearch} (the depth of exploration during the search) is also varied from 5 to 25, and the average \texttt{recall@10} is measured for each hybrid query set.

\paragraph{Experiment Results.} The results are presented in \autoref{fig:oracle}. In this figure, each set of hybrid queries is classified based on the distribution relationship between query vectors and base vectors, represented by circle, cross, and square for ID, POD, and OOD, respectively. 

For both datasets, the three sets of \textit{baseline hybrid queries} achieve the best performance, as the query vectors and base vectors belong to the same distribution, naturally classified as ID. For MNIST-8M, the remaining 6 sets of hybrid queries exhibit clear stratification and are classified as either POD or OOD, while for MTG, all 6 remaining sets have nearly identical query performance and are classified as POD. It is also observed that the relative query performance is consistent for both IVFFlat and HNSWFlat indices. This indicates that ID queries are consistently easier, while POD and OOD queries are relatively more difficult, regardless of the index type. Furthermore, compared to IVFFlat, HNSWFlat demonstrates significant speedup for ID queries but offers very little improvement for POD and OOD queries. This highlights the potential for further optimization of graph-based indices in handling queries across different distribution relationships. Notably, recent studies \cite{DBLP:journals/corr/abs-2211-12850, DBLP:journals/pvldb/ChenZHJW24} have made encouraging progress in this direction.

\paragraph{Qualitative Explanations.} To provide an initial overview of hybrid datasets and a qualitative understanding of distribution relationships between \textit{filtered subsets}, we visualize each dataset in a 2-dimensional space using Uniform Manifold Approximation and Projection (UMAP) \cite{DBLP:journals/corr/abs-1802-03426}. UMAP is a graph-based algorithm for dimensionality reduction. It first constructs a weighted k-neighbor graph and then computes a low-dimensional layout of this graph, capturing both the global structure, similar to techniques like PCA \cite{Hotelling1933} and Laplacian Eigenmaps \cite{DBLP:journals/neco/BelkinN03}, and the local structure, similar to techniques like t-SNE \cite{JMLR:v9:vandermaaten08a} and LargeVis \cite{DBLP:conf/www/TangLZM16}. This graph-based dimensionality reduction process, which is closely related to graph-based indices for vector similarity search, makes UMAP a suitable tool for visualization.

In \autoref{fig:umap}, we perform UMAP visualizations \footnote{\url{https://github.com/lmcinnes/umap/tree/release-0.5.7}} on the MNIST-8M and MTG datasets. Each \textit{filtered subset} has a 2-dimensional distribution. For MNIST-8M, we uniformly sample 100,000 data points. The visualization reveals 10 well-separated distributions of \textit{filtered subsets}, exhibiting clustering behavior, and the number of data points is balanced across \textit{filtered subsets}. For MTG, we use all its 40,274 data points. The visualization reveals 6 highly overlapping distributions of \textit{filtered subsets}, sharing a similar overall distribution, and the majority of data points are concentrated with 3 scalar values (\texttt{common}, \texttt{rare}, and \texttt{uncommon}).

The visualizations partially explain the experimental results shown in \autoref{fig:oracle}. For MNIST-8M, hybrid queries such as those between \texttt{4} and \texttt{9} are classified as ID, as their handwritten digits are visually similar, leading to closer image embeddings and a relatively large overlap between their distributions of \textit{filtered subsets}. In contrast, queries such as those between \texttt{4} and \texttt{7} are classified as OOD due to the significant visual differences in their handwritten digits, resulting in highly separated distributions of \textit{filtered subsets}. For MTG, all \textit{filtered subsets} share a similar overall distribution, resulting in nearly identical query performance across all 6 remaining sets of hybrid queries, which are classified as ID. However, the 2-dimensional nature of UMAP visualizations cannot fully capture the distribution relationships in high-dimensional space. For example, for MNIST-8M, the distributions of \textit{filtered subsets} for \texttt{7} and \texttt{9} exhibit some degree of overlap, but the performance for \texttt{9}-\texttt{7} is significantly worse than that for \texttt{7}-\texttt{9}, leading the former to be classified as OOD and the latter as POD.

\begin{figure*}[tbp]
  \centering
  \includegraphics[width=0.85\linewidth]{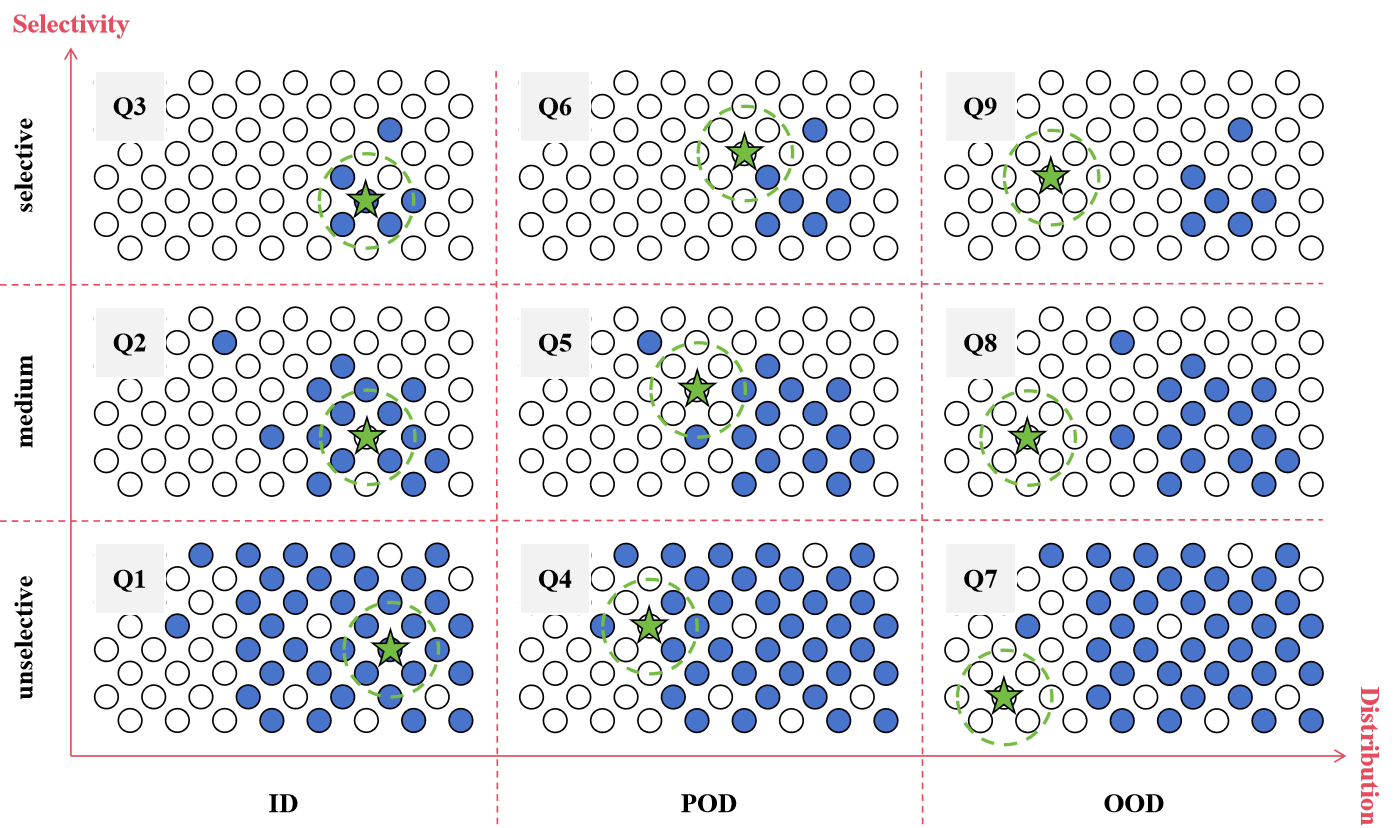}
  \caption{Hybrid query sets with varying levels of difficulty under the control of both the \textit{distribution} factor and the \textit{selectivity} factor. Each region represents a hybrid query set, where discs represent data points, base vectors are blue discs, and query vectors locate in a green dashed circular area centered on a green pentagram. A lower proportion of blue discs indicates higher \textit{selectivity}, and less overlap between the green and blue area indicates greater proximity to OOD.}
  \label{fig:query}
\end{figure*}

\paragraph{Quantitative Explanations.} To complement qualitative visualizations and provide a quantitative understanding of distribution relationships between \textit{filtered subsets}, we use the Mahalanobis distance \cite{Mahalanobis1936}, which has been employed in recent studies \cite{DBLP:journals/corr/abs-2211-12850, DBLP:journals/pvldb/ChenZHJW24} to quantify the OOD property in vector similarity search. The Mahalanobis distance measures the distance from a vector $\mathbf{v}$ to a \textit{filtered subset} $\mathcal{D}_{f_{s}}$, and is defined as $d_{M}(\mathbf{v}, \mathcal{D}_{f_{s}}) = \newline \sqrt{(\mathbf{v} - \mathbf{\bar{v}_{\mathcal{D}_{f_{s}}}})^T S_{\mathcal{D}_{f_{s}}}^{-1} (\mathbf{v} - \mathbf{\bar{v}_{\mathcal{D}_{f_{s}}}})}$, where $\mathbf{\bar{v}}_{\mathcal{D}_{f_{s}}}$ is the mean vector of $\mathcal{D}_{f_{s}}$ and $S_{\mathcal{D}_{f_{s}}}^{-1}$ is the inverse of the covariance matrix of $\mathcal{D}_{f_{s}}$.

In \autoref{fig:mahalanobis}, we calculate Mahalanobis distance histograms for all sets of hybrid queries in MNIST-8M and MTG. For a set of hybrid queries ``$s_{base} - s_{query}$'' mentioned above, we uniformly sample $\hat{\mathcal{D}}_{f_{s_{base}}}$ from $\mathcal{D}_{f_{s_{base}}}$ and $\hat{\mathcal{D}}_{f_{s_{query}}}$ from $\mathcal{D}_{f_{s_{query}}}$, each containing 5,000 data points. Notably, when $\hat{\mathcal{D}}_{f_{s_{base}}} = \hat{\mathcal{D}}_{f_{s_{query}}}$, the two sampled subsets are required to be non-intersecting. We then use an open-source library \footnote{\url{https://github.com/mosegui/mahalanobis}. We modify the original library to compute the pseudo-inverse of the covariance matrix instead of the regular inverse for robustness.} to compute the Mahalanobis distances in the Euclidean space. Each subgraph titled ``$base-query$'' shows both the orange histogram of Mahalanobis distances from each vector in $\hat{\mathcal{D}}_{f_{s_{base}}}$ to $\hat{\mathcal{D}}_{f_{s_{query}}}$, and the blue histogram of Mahalanobis distances from each vector in $\hat{\mathcal{D}}_{f_{s_{base}}}$ to $\hat{\mathcal{D}}_{f_{s_{base}}}$, illustrating the distribution shift between query vectors and base vectors when sampled from different \textit{filtered subsets}, compared to when they are sampled from the same \textit{filtered subset}.

The calculations also partially explain the experimental results shown in \autoref{fig:oracle}. For both datasets, in each diagonal subgraphs (from the top left to the bottom right), two histograms show complete overlap between orange and blue, which explains why \textit{baseline hybrid queries} achieve the highest performance and are classified as ID (\autoref{fig:oracle}). The off-diagonal subgraphs are asymmetric, as the Mahalanobis distance depends on the covariance matrix of the base vectors. This explains why \texttt{7}-\texttt{9} is POD but \texttt{9}-\texttt{7} is OOD (\autoref{fig:oracle}). In each off-diagonal subgraph, two histograms exhibit some degree of overlap. Notably, \texttt{4}-\texttt{9} and \texttt{7}-\texttt{9} in MNIST-8M and all the remaining pairs in MTG show near-complete overlap between orange and blue, which explains their relative high performance and are classified as POD (\autoref{fig:oracle}). However, the remaining subgraphs in MNIST-8M cannot be classified as POD or OOD based solely on the size of the overlapping region. For example, while the overlap size suggests that \texttt{4}-\texttt{7} $>$ \texttt{9}-\texttt{7} $>$ \texttt{9}-\texttt{4}, these pairs are classified as OOD, POD, and OOD, respectively (\autoref{fig:oracle}). 

The above analysis suggests that the Mahalanobis distance is still imperfect. Its success in distinguishing OOD queries in existing studies \cite{DBLP:journals/corr/abs-2211-12850, DBLP:journals/pvldb/ChenZHJW24} can be largely attributed to the fact that the base vectors and query vectors are sampled from different modalities (e.g., text and image embeddings generated by CLIP \cite{DBLP:conf/icml/RadfordKHRGASAM21}), which are inherently far apart and have almost no overlap, making them easy to distinguish. However, in the context of hybrid queries, the distinguishing ability of Mahalanobis distance is insufficient, as it fails to completely classify query properties (POD or OOD) based on the degree of histogram overlap. This highlights the need to develop a more precise metric, which can better support the design of the set of hybrid queries with desired distribution relationship between base vectors and query vectors. Notably, a recent study \cite{DBLP:journals/pvldb/ChenZHJW24} explored the Wasserstein distance as an alternative, but it suffers from symmetry, meaning that the distances between \texttt{x}-\texttt{y} and \texttt{y}-\texttt{x} are the same, making it less effective than Mahalanobis distance in explaining query difficulty.

\subsection{Towards More Comprehensive Evaluation of FANNS Algorithms}
\label{subsec:Towards More Comprehensive Evaluation of FANNS Algorithms}

While the \textit{selectivity} factor measures the proportion of data points excluded by the scalar filter, the \textit{distribution} factor provides insights into the relationship between distributions of base vectors and query vectors. As discussed above, both factors contribute to the hybrid query difficulty. Combining these two factors, we propose a schema towards more comprehensive evaluation of FANNS algorithms, in which each factor serves as a dimension to partition the space of hybrid query sets with varying levels of difficulty.

\autoref{fig:query} illustrates the space partitioned by these two factors, \textit{distribution} and \textit{selectivity}. There are 9 sets of hybrid queries (\textbf{Q1}-\textbf{Q9}) across 9 regions, with each set representing a combination of a specific \textit{distribution} and \textit{selectivity}. In each region, discs represent data points, the base vectors are discs with blue color, and the query vectors are enclosed within a green dashed circular area centered on a green pentagram. A lower proportion of blue discs in a region indicates higher \textit{selectivity}, and less overlap between the green and blue area indicates greater proximity to OOD.

Although the \textit{selectivity} can be easily controlled by adjusting the range of scalar values, there is currently no straightforward method for controlling the \textit{distribution}. As discussed in \autoref{subsec:Impact of Distribution Factor: Experiments and Explanations}, an effective metric for quantifying the distribution relationship between base vectors and query vectors is still lacking. Consequently, the practical realization of our proposed schema depends on the development of a more suitable metric for measuring the \textit{distribution} factor. We view it as an important research direction coming out of this survey.

\section{Open Questions and Research Directions}
\label{sec:Open Questions and Research Directions}

Aside from the above-mentioned research direction, this section discusses open questions and possible research directions in the field of FANNS, progressing from the internal structures of its indices, to the workload-aware optimization of its algorithms, and ultimately to its system-level solutions.

\subsection{Innovation in Index Structures}
\label{subsec:Innovation in Index Structures}

Designing efficient FANNS indices through the integration of traditional data structures is a notable trend.

For example, NHQ (\textbf{A7}) leverages the \textit{prefix tree} \cite{ACM:10.1145/3698822}, iRangeGraph (\textbf{A11}) and WST (\textbf{A17}) utilize the \textit{segment trees} \cite{DBLP:conf/icml/EngelsLYDS24, DBLP:journals/corr/abs-2409-02571}, and HQI (\textbf{A14}) employs the \textit{qd-tree} \cite{DBLP:journals/pacmmod/MohoneyPCMIMPR23}. These integrations demonstrate that traditional data structures can play a crucial role in building more efficient FANNS indices.

However, existing FANNS indices integrated with traditional data structures are tightly coupled with specific assumptions, such as a \textit{simplified equality scalar filter} or a \textit{simplified range scalar filter} (\autoref{subsec:Definition of Hybrid Query}). Therefore, a key open question remains: how to integrate or design data structures that can support efficient FANNS indexing under more general scalar filters.

\subsection{Optimization via Realistic Workloads}
\label{subsec:Optimization via Realistic Workloads}

Optimizing dedicated FANNS algorithms based on realistic workloads is another important research direction.

For instance, CAPS (\textbf{A6}) partitions clusters of the IVF index according to the power-law distribution of scalar values \cite{DBLP:journals/corr/abs-2308-15014}, Milvus (\textbf{A13}) performs pre-indexing dataset partitioning based on frequently queried scalar values \cite{DBLP:conf/sigmod/WangYGJXLWGLXYY21}, and HQI (\textbf{A14}) benefits from workloads that exhibit scalar filter stability \cite{DBLP:journals/pacmmod/MohoneyPCMIMPR23}. 

The primary challenge here is how to effectively identify and quantify workload characteristics in specific application scenarios, and to perform targeted optimizations based on these characteristics.

\subsection{Combination of Multiple Algorithms}
\label{subsec:Combination of Multiple Algorithms}

Combining multiple FANNS algorithms and dynamically select the most suitable one for a given hybrid query is a promising direction at the system level. This strategy enables the system to exploit the strengths of different FANNS algorithms under varying conditions.

Several studies have explored this strategy by estimating \textit{selectivity} for algorithm selection, typically using Pre-Filtering Algorithm Family (\textbf{A12}) for high \textit{selectivity} and pre-built FANNS indices for moderate or low \textit{selectivity}. For example, database systems such as ADBV \cite{DBLP:journals/pvldb/WeiWWLZ0C20}, Milvus \cite{DBLP:conf/sigmod/WangYGJXLWGLXYY21}, and VBase \cite{DBLP:conf/osdi/ZhangXCSXCCH00Y23} utilize cost models to estimate \textit{selectivity} and dynamically switch between FANNS algorithms. In the case of ACORN (\textbf{A4}), the algorithm shifts to Pre-Filtering Algorithm Family (\textbf{A12}) when \textit{selectivity} is high \cite{DBLP:journals/pacmmod/PatelKGZ24}. 

Remaining challenges include identifying and estimating more workload-aware metrics beyond \textit{selectivity} to guide algorithm selection, and optimizing storage efficiency when multiple indices coexist.

\section{Conclusion}

In this paper, we formally defined the FANNS problem, covering the hybrid dataset, the hybrid query, and key evaluation metrics. We then proposed a pruning-focused framework to classify and summarize existing FANNS algorithms. Next, we reviewed existing hybrid datasets, outlining their construction strategies and detailing their contents. We also endeavored to understand the difficulty of hybrid queries by incorporating \textit{distribution} in addition to \textit{selectivity}, providing insights through qualitative visualizations and quantitative measurements, and proposed a schema towards more comprehensive evaluation of FANNS algorithms. Finally, we highlighted open questions and possible research directions. Going forward, we will try to develop a benchmark for enabling more comprehensive evaluation of FANNS algorithms.


\begin{acknowledgements}
This paper was supported by National Natural Science Foundation of China (U24A20232), and Key Technology Research and Development Program of Shandong Province (2024CXGC010113).
\end{acknowledgements}

\bibliographystyle{spbasic} 
\bibliography{reference} 

\end{document}